\documentclass[
superscriptaddress,citeautoscript,floatfix,longbibliography,bibnotes]{revtex4}
\usepackage{times,bm,siunitx,amsmath,amsfonts,amsthm,amssymb,amsbsy,braket,graphicx,orcidlink}
\usepackage[utf8]{inputenc}
\usepackage[T1]{fontenc}
\usepackage[capitalize]{cleveref}
\renewcommand{\Re}{\mathrm{Re}}
\renewcommand{\Im}{\mathrm{Im}}
\DeclareMathOperator{\sgn}{sgn}
\hypersetup{colorlinks=true,urlcolor=blue,linkcolor=blue,citecolor=blue}
\usepackage{microtype}
\usepackage{multirow}

\begin{document}
\title{
Nonhermitian topological zero modes at smooth domain walls:
Exact solutions 
}
\author{Pasquale Marra \orcidlink{0000-0002-9545-3314}}
\email{pasquale.marra@keio.jp}
\affiliation{Department of Engineering and Applied Sciences, Sophia University, 7-1 Kioi-cho, Chiyoda-ku, Tokyo 102-8554, Japan}
\affiliation{Department of Physics, and Research and Education Center for Natural Sciences, Keio University, 4-1-1 Hiyoshi, Yokohama, Kanagawa, 223-8521, Japan}
\affiliation{Graduate School of Informatics, Nagoya University, Furo-cho, Chikusa-Ku, Nagoya, 464-8601, Japan}
\author{Angela Nigro \orcidlink{0000-0001-8326-5781}}
\email{anigro@unisa.it}
\affiliation{Dipartimento di Fisica ``E. R. Caianiello'', Università degli Studi di Salerno, 84084 Fisciano (Salerno), Italy}
\date{\today}

\begin{abstract}
The bulk-boundary correspondence predicts the existence of boundary modes localized at the edges of topologically nontrivial systems.
The wavefunctions of hermitian boundary modes can be obtained as the eigenmodes of a modified Jackiw-Rebbi equation.
The bulk-boundary correspondence has also been extended to nonhermitian systems, which describe physical phenomena such as gain and loss in open and non-equilibrium systems.
Nonhermitian energy spectra can be complex-valued and exhibit point gaps or line gaps in the complex plane, whether the gaps can be continuously deformed into points or lines, respectively.
Specifically, line-gapped nonhermitian systems can be continuously deformed into hermitian gapped spectra.
Here, we find the analytical form of the wavefunctions of nonhermitian boundary modes with zero energy localized at smooth domain boundaries between topologically distinct phases by solving the generalized Jackiw-Rebbi equation in the nonhermitian regime.
Moreover, we unveil a universal relation between the scalar fields and the decay rate and oscillation wavelength of the boundary modes.
This relation quantifies the bulk-boundary correspondence in nonhermitian line-gapped systems through physical quantities that are experimentally measurable.
Furthermore, this relation is not affected by the specific spatial variations of the scalar fields.
These results offer new insights into the localization properties of boundary modes in nonhermitian and topologically nontrivial states of matter. 
\end{abstract}
\maketitle

\section{Introduction}

The bulk-boundary correspondence determines the presence and number of localized modes in relation to the topological invariants in topologically nontrivial condensed matter systems~\cite{hatsugai_chern_1993,ryu_topological_2002,teo_topological_2010}, such as topological insulators and superconductors~\cite{schnyder_classification_2009,hasan_colloquium:_2010,qi_topological_2011,shen_topological_2011,shen_topological_2017}.
However, the bulk-boundary correspondence cannot predict the physical properties of the localized modes, such as their localization length, the presence or absence of oscillations in the wavefunction, and other local properties.
Generalizations of the bulk-boundary correspondence apply to nonhermitian systems~\cite{ashida_non-hermitian_2020,okuma_non-hermitian_2023} as well.
Nonhermitian Hamiltonians naturally describe a variety of physical systems, such as open systems in the presence of gain and loss mechanisms, or dissipation, and exhibit the presence of complex energy spectra~\cite{shen_topological_2018,gong_topological_2018,kawabata_symmetry_2019,ashida_non-hermitian_2020,okuma_non-hermitian_2023,schindler_hermitian_2023,nakamura_bulk-boundary_2024} in several different condensed matters and metamaterials~\cite{gao_observation_2015,ghatak_observation_2020,zhang_observation_2021,su_direct_2021,fleckenstein_non-hermitian_2022,lin_observation_2022,liu_experimental_2023,wang_non-hermitian_2023,ochkan_non-hermitian_2024}.
A fundamental distinction in this case is between spectra with line gaps and point gaps~\cite{shen_topological_2018,gong_topological_2018,kawabata_symmetry_2019}.
Point gaps describe energy eigenvalues that do not cross a reference point (usually the origin at zero). 
In contrast, line gaps describe energy eigenvalues that do not cross a reference line in the complex plane.
Any nonhermitian Hamiltonian with line gaps can be adiabatically deformed into a corresponding hermitian Hamiltonian with conventional energy gaps.
Hence, the bulk-boundary correspondence and the definition of topological invariants can be directly extended in this case.
Hermitian Hamiltonians with nontrivial topology are naturally described by a modified Jackiw-Rebbi equation~\cite{shen_topological_2011,shen_topological_2017}, a spinorial second-order linear differential equation.
This generalizes the original Jackiw-Rebbi equation~\cite{jackiw_solitons_1976}, which is instead of first-order.
Henceforth, nonhermitian Hamiltonians can be described by a nonhermitian version of the modified Jackiw-Rebbi equation with complex coefficients.

In recent works, we found the analytical wavefunctions of the zero energy modes of the modified Jackiw-Rebbi equation localized at smooth domain walls between topologically inequivalent phases in the hermitian regime~\cite{marra_topological_2024} and extended some of these findings to the nonhermitian regime~\cite{marra_zero_2025}.
In this work, we derive the analytical form of the wavefunctions of zero modes in nonhermitian line-gapped systems described by a modified Jackiw-Rebbi equation with space-dependent mass and velocity terms, allowing for complex coefficients.
The space dependency describes the case of a smooth domain wall, i.e., the case where the scalar fields are nonuniform within a region of finite length $w$, which we call a \emph{smooth} domain wall, and become uniform at larger distances $x \gg w$.
This allows us to derive the bulk-boundary correspondence for nonhermitian line-gapped systems by directly establishing the relation between nonhermitian topological invariants and the asymptotic decay rate of the zero modes.
We thus analyze several properties of these zero modes, such as the localization length $\xi$, oscillation wavelength $\lambda$, conditions for the reality of the wavefunction, local properties, and local topological invariant, and show specific examples of nonhermitian zero modes corresponding to different regimes.
This allows us to classify zero modes in different categories:
Featureless zero modes in the limit $w\to0$;
Non-featureless zero modes with localization length and oscillation wavelength are larger than the width of the smooth domain wall $\xi,\lambda>w>0$;
Non-featureless zero modes where the width of the smooth domain wall is larger than the localization length and oscillation wavelength $w>\xi,\lambda>0$.
Featureless zero modes are described only by their localization lengths and oscillation wavelengths on the two sides of the smooth domain wall:
For this reason, we call these zero modes with "no hair", since they can be completely described by only two numbers (in analogy with the terminology used for black holes).
By contrast, we call non-featureless zero modes as zero modes with "hair".
We can further distinguish zero modes with "short hair" for $\xi,\lambda>w>0$, which appear featureless at long length scales, and with "long hair" for $w>\xi,\lambda>0$, which appear non-featureless at all length scales.
We also discuss possible experimental signatures of the bulk-boundary correspondence by unveiling a universal relation between the localization lengths, oscillation wavelengths, and bulk physical quantities. 

\section{Exact wavefunctions of the zero energy modes}

\subsection{Modified Jackiw-Rebbi equation in the nonhermitian regime}

\begin{figure}[tbp]
 \centering
 \includegraphics[width=.5\textwidth]{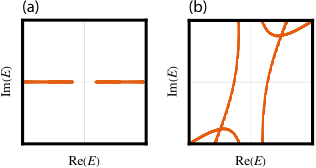}
\caption{
Energy spectra of the generalized Jackiw-Rebbi equation with uniform fields $v(x)=v$, $m(x)=m$ in \cref{eq:Hk}.
(a) Hermitian case with $v,m\in\mathbb{R}$ corresponding to a real energy spectrum with an energy gap.
(b) nonhermitian case with $v,m\notin\mathbb{R}$ corresponding to a complex energy spectrum with a line gap.
}
 \label{fig:linegap}
\end{figure}

The Jackiw-Rebbi equation~\cite{jackiw_solitons_1976} is a Dirac equation where the mass term is space-dependent.
Here, we consider a generalization of the Jackiw-Rebbi equation, where both the mass and Dirac velocity are space-dependent
\begin{equation}
	\left[\left(\eta p^2+{m}(x)\right)\sigma_z
	+
 {2{v}}(x)
 p\,\sigma_y
 \right]\Psi(x)=E \Psi(x).
\label{eq:JR}
\end{equation}
with $\Psi(x)$ a two-component spinor describing the fermionic field, $\sigma_{xyz}$ the Pauli matrices, $p=-\mathrm{i}\partial$ the momentum operator, and $\eta\ge0$.
Notice that the equation above differs from the original Jackiw-Rebbi equation by the addition of a quadratic term in the momentum.
We consider here the nonhermitian case where the fields are complex-valued ${m}(x),{v}(x)\in\mathbb{C}$.
This equation can be interchangeably referred to as a modified Jackiw-Rebbi equation or a modified Dirac equation with space-dependent mass and Dirac velocity.
The hermitian case ${m}(x),{v}(x)\in\mathbb{R}$ describes topological insulators and superconductors and has been studied at length in our previous work Ref.~\cite{marra_topological_2024}. 
The nonhermitian case $\Im({m}(x))\neq0$ describes the presence of a nonhermitian potential $V(x)={m}(x)$. 
Examples of such systems are topological insulators perturbed by
a space-dependent imaginary potential, describing gain and loss dynamics.
In particular, smoothly varying, space-dependent gain and loss terms have been implemented in photonic lattices in several experiments~\cite{wimmer_observation_2015,bandres_topological_2018,zhao_non-hermitian_2019,wang_hybridization_2025}.
On the other hand, the nonhermitian case $\Im({v}(x))\neq0$ describes, e.g., imbalanced superconducting pairing corresponding to dissipation effects on Cooper pairs.
Examples of such systems are Kitaev chains with imbalanced superconducting pairing~\cite{li_topological_2018,zhao_anomalous_2021,zhao_defective_2021} and other nonhermitian superconducting states~\cite{ghatak_theory_2018}.
Note that the term $\propto2v(x)p\sigma_y$ does not describe the familiar case of a gauge-invariant hermitian $p$-wave superconducting pairing with arbitrary complex phase $\propto\Re(\Delta(x))p\sigma_y+\Im(\Delta(x))p\sigma_x=\Big(\begin{smallmatrix}0&-\Delta^*(x)\partial\\\Delta(x)\partial&0 \end{smallmatrix}\Big)$ but an imbalanced nonhermitian pairing $\propto\Re(\Delta(x))p\sigma_y+\Im(\Delta(x))p\sigma_y=\Big(\begin{smallmatrix}0&-\Delta(x)\partial\\\Delta(x)\partial&0 \end{smallmatrix}\Big)$ (we remind the reader that differentiation is an antihermitian operation). 
In the hermitian case, our model describes zero modes in one dimension with time-reversal, charge conjugation, and chiral symmetry, corresponding to the Altland-Zirnbauer class BDI (see Ref.~\cite{marra_topological_2024}).
In the nonhermitian case, this corresponds to nonhermitian topological phases with real line gaps (see below) in one dimension in the Altland-Zirnbauer class BDI.
However, since all nonhermitian topological phases can also be described in the continuum by a generalized Jackiw-Rebbi nonhermitian Hamiltonian~\cite{shen_topological_2018}, these findings can in principle be extended to other nonhermitian topological phases with real line gaps, imaginary line gaps, and point gaps in all Altland-Zirnbauer classes in one or higher dimensions~\cite{kawabata_symmetry_2019}.

We will consider the zero modes $E=0$ that are normalizable and satisfy the Dirichlet boundary conditions on infinite or semi-infinite intervals.
Zero energy modes of \cref{eq:JR} are eigenstates of $\sigma_x$, i.e., $\sigma_x\Psi(x)=s\Psi(x)$, 
where the eigenvalue $s=\pm1$ is referred to as the pseudospin.
Hence, the zero modes are given by
$\Psi(x)\propto
\tiny\begin{pmatrix} 
1 \\ 
s
\end{pmatrix}
\varphi(x)$,
with $s=\pm1$ and where $\varphi(x)$ satisfies 
\begin{equation}\label{eq:diffeq}
\eta\varphi''(x)
+{2s{v}}(x) \varphi'(x)
-
{m}(x)
\varphi(x)
=0.
\end{equation}
Hereafter we take $\eta=1$, without loss of generality, since the scalar fields can be always redefined as ${v}(x)\to {v}(x)/\eta$, ${m}(x)\to {m}(x)/\eta$.
If $\Im({m}(x))\neq0$ or $\Im({v}(x))\neq0$, the equation is not real and does not have real solutions in the general case.
We recall that real wavefunctions are the hallmark of Majorana zero modes.
In other words, the zero modes of a nonhermitian system are not necessarily Majorana zero modes.

\subsection{Line gap nonhermitian topology}

Assuming uniform fields $v(x)=v$, $m(x)=m$, the Hamiltonian in \cref{eq:JR} becomes diagonal in the momentum eigenfunctions and can be written as
\begin{equation}
	\left(\left(k^2+{m}\right)\sigma_z
	+
 {2{v}}
 k\,\sigma_y
 \right)\Psi(k)=E \Psi(k),
\label{eq:Hk}
\end{equation}
with an energy spectra given by $E(k)^2=(k^2+m)^2+4v^2k^2$.
In the nonhermitian case, the energy spectra are generally complex.
In nonhermitian systems~\cite{shen_topological_2018,gong_topological_2018,kawabata_symmetry_2019}, energy spectra can exhibit either point gaps, where the energy eigenvalues do not cross a specific point in the complex plane, or line gaps, where the energy eigenvalues do not cross a specific line in the complex plane.
Line gaps can be further classified as real line gaps, where the energy levels cannot cross the imaginary axis, and imaginary line gaps, where the energy levels cannot cross the real axis.
Nonhermitian spectra with line gaps can always be adiabatically deformed to hermitian and gapped spectra.
The nonhermitian complex energy spectra of \cref{eq:Hk} shows a 
real 
line gap, which is the line in the complex plane with $\Im(E(k))=0$, and can be deformed to hermitian real energy spectra with an energy gap by taking the limits $\Im(v)\to0$, $\Im(m)\to0$, as shown in \cref{fig:linegap}.
The line gap closes when $E(k)=0$ for some momenta $k\in\mathbb{R}$.
Hence, the closing of the line gap is determined if the condition $E(k)=0$ is satisfied for any $k\in\mathbb{R}$, i.e., if any of the solutions of the algebraic equation $(k^2+m)^2+4v^2k^2=0$, which are given by 
$k=\pm\sqrt{-\left(\sqrt{m+v^2}\pm\sqrt{v^2}\right)^2}$,
is real.
This condition is satisfied only if $\Re\left(\sqrt{m+v^2}\pm\sqrt{v^2}\right)=0$.
Hence, the line gap of the complex energy spectra closes when the quantity
\begin{equation}
M=|\Re(\sqrt{{v}^2+{m}})|-|\Re({{v}})|,
\end{equation}
becomes zero.
The closing of the line gap separates topologically nonequivalent phases with $M>0$ and $M<0$.
In the hermitian limit, one has that $M=0$ if and only if $m=0$, and therefore, the condition above simply reduces to the condition $m=0$.
To determine the topological invariant $W$ of the gapped spectra, we recall that in the hermitian case, the topologically trivial phase with $W=0$ corresponds to $m>0$, and the topologically nontrivial phases with $W\neq0$ correspond to $m<0$ with $W=\sgn(v)$.
Since the nonhermitian topological invariant must coincide with the hermitian topological invariant in the limit $\Im(v)\to0$, $\Im(m)\to0$, one can conclude that
\begin{equation}\label{eq:TI0}
W=
\begin{cases}
\sgn(\Re({v})), & \text{ if } \sgn({M})=-1 \text{ or } 0,
\\
0 & \text{ if } \sgn({M})=1.
\end{cases}
\end{equation}

\subsection{Exact wavefunctions}

In this Section, we quickly derive the general solutions of the Jackiw-Rebbi equation in the nonhermitian regime by directly generalizing the solutions in the hermitian regime found in our previous work~\cite{marra_zero_2025}.
For a detailed derivation of these solutions, see \cref{app:formal}.
We assume that the fields become constant at large distances $x\to\pm\infty$ and that they approach their asymptotic values exponentially as
$|{m}(x\to\mp\infty)-{m}_{L,R}|\sim{e}^{- |x|/w}$,
$|{v}(x\to\mp\infty)-{v}_{L,R}|\sim{e}^{- |x|/w}$.
Here, the characteristic length $w>0$ measures the width of the smooth domain wall.
Under these very general assumptions, using the substitution $y(x)=(1+\tanh{(x/2w)})/2$, we map the whole real line $(-\infty,\infty)$ into the finite segment $(0,1)$.
This allows us to find the eigenmodes of the modified Jackiw-Rebbi equation given by $A_1\varphi_1^s(x)+A_2\varphi_2^s(x)$ with
\begin{align}
 \varphi_{1,2}^s(x)&=
y(x)^{{w\alpha_L^\pm}}(1-y(x))^{{w\alpha_R^\mp}} 
F_{1,2}(y(x))
\nonumber\\&=
 \frac{{e}^{({ \alpha_L^\pm}-{\alpha_R^\mp})x/2} }
 {\left(2\cosh{(x/2w)}\right)^{{w(\alpha_L^\pm}+{\alpha_R^\mp})}}
 F_{1,2}(x)
 ,
\label{eq:2generalsolutionsMAIN}
\end{align}
for a given pseudospin $s=\pm1$, and with $\alpha_L^\pm=-s{{v}}_L\pm{q}_{L}$, 
$\alpha_R^\pm=s{{v}}_R\pm{q}_{R}$,
where ${q}_{L,R}=\sqrt{{v}_{L,R}^2+{m}_{L,R}}$.
A fundamental property of the exponents $\alpha_{L,R}^\pm$ is that they only depend on the asymptotic values of the scalar fields $|x|\gg w$.
The functions $F_{1,2}$ are bounded or diverge polynomially for $x\to\pm\infty$.
The detailed derivation of these solutions is given in \cref{app:formal}.
The asymptotic behavior of these eigenmodes is
\begin{subequations}\label{eq:decayMAIN}
\begin{align}
\varphi_{1,2}^s(x\to-\infty)&\sim
{e}^{- \alpha_L^\pm |x|}={e}^{- \mu_L^\pm |x|}{e}^{- \mathrm{i}\kappa_L^\pm |x|},
\\
\varphi_{1,2}^s(x\to+\infty)&\sim{e}^{- \alpha_R^\mp |x|}={e}^{- \mu_R^\mp |x|}{e}^{- \mathrm{i}\kappa_R^\mp |x|}.
\end{align}
\end{subequations}
Here, $\mu_{L,R}^\pm=\Re(\alpha_{L,R}^\pm)$ are the decay rates on the opposite sides of the smooth domain wall, and
$\xi_{L,R}^\pm=1/\Re(\alpha_{L,R}^\pm)$ the corresponding decay lengths;
$\kappa_{L,R}^\pm=\Im(\alpha_{L,R}^\pm)$ are the momenta describing the oscillatory behavior of the eigenmodes on the opposite sides of the smooth domain wall, and
$\lambda_{L,R}^\pm=2\pi/\Im(\alpha_{L,R}^\pm)$ the corresponding wavelengths.
Obviously $\mu_{L,R}^+=\mu_{L,R}^-=\mp s \Re({v}_{L,R})$ if $\Re({q}_{L,R})=0$ and $\kappa_{L,R}^+=-\kappa_{L,R}^-=\Im({q}_{L,R})$ if $\Im({v}_{L,R})=0$.

We can further assume that the scalar fields can be expanded in powers of $y(x)$.
Specifically, we assume ${v}(x)={v}_0+{v}_1 y(x)$ (up to the first order) and ${m}(x)={m}_0+{m}_1 y(x)+{m}_2 y(x)^2$ (up to the second order).
Here, the coefficients of the expansion can be related to the values of the fields at large distances as $v_0=v_L$, ${v}_1={v}_R-{v}_L$, and $m_0=m_L$, $m_1+m_2={m}_R-{m}_L$, ${m}_2= 2 ({m}_L + {m}_R) - 4 {m}_D$ and to the value of the mass at the center of the domain wall ${m}_D={m}(0)$.
With these further assumptions, we get that the eigenmodes are given by \cref{eq:2generalsolutionsMAIN} with
\begin{align}
F_{1,2}(x)=&
{{{\,}_2F_1}\left(a_{1,2},b_{1,2},c_{1,2},\tfrac{1}{{e}^{\mp \, x/w}+1}\right)}
,
\label{eq:2generalsolutionshyperMAIN}
\end{align}
where
${{\,}_2F_1}(a,b,c,y)$ is the hypergeometric function and
\begin{subequations}\label{eq:ab1ab2MAIN}
\begin{align} 
a_{1,2}&=\pm w\left({{q}_L}-{{q}_R}\right) +\tfrac12+\tfrac12{\sqrt{\left(2ws{v}_1- 1\right)^2+4w^2{m}_2}}
\\
b_{1,2}&=\pm w\left({{q}_L}-{{q}_R}\right) +\tfrac12-\tfrac12{\sqrt{\left(2ws{v}_1- 1\right)^2+4w^2{m}_2}}
\\
c_{1,2}&=1+2w {q}_{L,R},
\end{align}
\end{subequations}
assuming $c_{1,2}-a_{1,2},c_{1,2}-b_{1,2}\notin\mathbb{Z}_\leq$ (we denote by $\mathbb{Z}_\leq$ the set of nonpositive integers).
The detailed derivation of this solution is given in \cref{app:exact}.
These solutions generalize to the nonhermitian case the solutions derived in Ref.~\cite{marra_topological_2024}.
The asymptotic behavior of the functions $F_{1,2}$ is summarized in \cref{tab:JRasymptotics}.
If either $c_{1}-a_{1}=c_{2}-a_{2}\in\mathbb{Z}_\leq$ or $c_{1}-b_{1}=c_{2}-b_{2}\in\mathbb{Z}_\leq$, the solutions of the hypergeometric equation become more complicated, since some of the hypergeometric functions become linearly dependent or ill-behaved for $x\to\pm\infty$.
We will not address these cases explicitly in this work.

\begin{table*}[t]
 \centering
 \begin{tabular}{@{} lll @{}}
\hline\\
&\qquad\qquad&
$F_{1,2}(x\to\mp\infty)=1$
\\[-1mm]\\\hline\\[-1mm]
$a_{1,2}\in\mathbb{Z}_\leq$ or $b_{1,2}\in\mathbb{Z}_\leq$
&\qquad&
$F_{1,2}(x\to\pm\infty)=
\frac{\Gamma(c_{1,2})\Gamma(c_{1,2}-a_{1,2}-b_{1,2})}{\Gamma(c_{1,2}-b_{1,2})\Gamma(c_{1,2}-a_{1,2})}\neq0$
\\[-1mm]\\\hline\\[-1mm]
$\Re({q}_{R,L})>0$,
$a_{1,2},b_{1,2}\notin\mathbb{Z}_\leq$ 
&\quad&
$F_{1,2}(x\to\pm\infty)=
\frac{\Gamma(c_{1,2})\Gamma(2w{q}_{R,L})}{\Gamma(c_{1,2}-a_{1,2})\Gamma(c_{1,2}-b_{1,2})}\neq0
$
\\[-1mm]\\\hline\\[-1mm]
$\Re({q}_{R,L})=0$, ${q}_{R,L}\neq0$,
$a_{1,2},b_{1,2}\notin\mathbb{Z}_\leq$ 
&\quad&
$F_{1,2}(x\to\pm\infty)
\sim
{e}^{\mp \mathrm{i}\Im(2{q}_{R,L}) x}
\frac{\Gamma(c_{1,2})\Gamma(-2w{q}_{R,L})}{\Gamma(a_{1,2})\Gamma(b_{1,2})}
+
\frac{\Gamma(c_{1,2})\Gamma(2w{q}_{R,L})}{\Gamma(c_{1,2}-a_{1,2})\Gamma(c_{1,2}-b_{1,2})}$
\\[-1mm]\\\hline\\[-1mm]
${q}_{R,L}=0$,
$a_{1,2},b_{1,2}\notin\mathbb{Z}_\leq$ 
&\quad&
$F_{1,2}(x\to\pm\infty)\sim
\pm (x/w)
\frac{\Gamma(c_{1,2})}{\Gamma(a_{1,2})\Gamma(b_{1,2})}$
\\[-1mm]\\\hline\\[-1mm]
 \end{tabular}
\caption{
The asymptotic behavior of the functions $F_{1,2}$.
We assume $c_{1,2}-a_{1,2},c_{1,2}-b_{1,2}\notin\mathbb{Z}_\leq$.
In the case ${m}_{L,R},{v}_{L,R}\in\mathbb{R}$
one has
$\Re({q}_{L,R})>0$ if ${{v}}_{L,R}^2+{m}_{L,R}>0$,
and
$\Re({q}_{L,R})=0$ if ${{v}}_{L,R}^2+{m}_{L,R}\le0$.
}
 \label{tab:JRasymptotics}
\end{table*}

The eigenmodes in the case where one or both of the scalar fields are uniform or for some special symmetric cases such as when $|{v}_{L,R}|={v}$ or ${m}_{L,R}={m}$ can be found in Table I of Ref.~\cite{marra_topological_2024}.
Since the functions $F_{1,2}$ are bounded or diverge polynomially, as summarized in \cref{tab:JRasymptotics}, the exponents $\alpha_L^\pm$ and $\alpha_R^\pm$ uniquely determine the asymptotic behavior of the eigenmodes.
Hence, the existence of zero modes, i.e., particular solutions that are normalizable and satisfy the boundary conditions on a given interval, is uniquely and solely determined by the exponents $\alpha_L^\pm$ and $\alpha_R^\pm$.
Since these exponents only depend on $v_{L,R}$, $m_{L,R}$, one can infer that the existence and number of zero energy modes and their pseudospin $s$ is univocally determined by the values assumed by the scalar fields at large distances.

\section{Nonhermitian bulk-boundary correspondence}

\subsection{Topological invariant}

As we have seen, the general solutions of the Jackiw-Rebbi equation in the nonhermitian regime are a direct generalization of the solutions in the hermitian regime obtained in our previous work~\cite{marra_zero_2025}. 
However, determining the existence and number of zero modes in the nonhermitian regime, i.e., determining the existence and number of particular solutions satisfying the boundary conditions, is not as straightforward.

In the hermitian regime, the bulk-boundary correspondence~\cite{hatsugai_chern_1993,ryu_topological_2002,teo_topological_2010} determines a relation between topologically protected modes at the boundary between two topologically inequivalent phases.
For uniform and hermitian fields $m(x)= m\in\mathbb{R}$, $v(x)= v\in\mathbb{R}$, the topological invariant $W\in\mathbb{Z}_2$ is $W=\sgn(v)$ (nontrivial) for $ m\le0$ and $W=0$ (trivial) otherwise~\cite{kitaev_unpaired_2001}.
Hence, assuming uniform fields at the asymptotes $x\to\pm\infty$, the phase on the left $x\ll -w$ has topological invariant $W_L=\sgn( v_L)$ if $ m_L\le0$ and $v_L\neq0$, and $W_L=0$ otherwise.
On the other hand, the phase on the right $x\gg w$ gives $W_R=\sgn( v_R)$ if $ m_R\le0$ and $v_R\neq0$, and $W_R=0$ otherwise (see Refs.~\cite{kitaev_unpaired_2001,shen_topological_2011}).
The difference between the topological invariants on the opposite sides of the boundary $\Delta W=|W_L-W_R|$ gives the number of topologically protected zero-energy modes localized at the boundary. 

Here, we will generalize the bulk-boundary correspondence to nonhermitian systems, particularly to nonhermitian systems described by nonhermitian regime of the modified Jackiw-Rebbi equation.
These nonhermitian systems can be seen as an adiabatical "deformation" of the conventional hermitian systems.
Hence, the bulk gap of the hermitian Hamiltonian is transformed into a line gap in the nonhermitian case~\cite{schindler_hermitian_2023}.
We will define here a topological invariant for these nonhermitian Hamiltonians by an adiabatic deformation of the topological invariant in the hermitian case, justified by our analysis of the decay rates of the localized boundary modes of the modified Jackiw-Rebbi equation in \cref{eq:JR}.

In agreement with our previous considerations on the line gap nonhermitian topology and the definition of bulk topological invariant in \cref{eq:TI0}, we define
\begin{subequations}\label{eq:determinants}
\begin{align}
{M}_{L,R}=&
|\Re(\sqrt{{v}_{L,R}^2+{m}_{L,R}})|-|\Re({{v}}_{L,R})|=
\nonumber\\&
|\Re({q}_{L,R})|-|\Re({{v}}_{L,R})|
,
\\
{K}_{L,R}=&
|\Im(\sqrt{{v}_{L,R}^2+{m}_{L,R}})|+|\Im({{v}}_{L,R})|=
\nonumber\\&
|\Im({q}_{L,R})|+|\Im({{v}}_{L,R})|
,
\\
\label{eq:TI}
W_{L,R}=&
\begin{cases}
\sgn(\Re({v}_{L,R})), & \text{ if } \sgn({M}_{L,R})=-1 \text{ or } 0,
\\
0 & \text{ if } \sgn({M}_{L,R})=1,
\end{cases}
\end{align}
\end{subequations}
where we use the common convention $\sgn(0)=0$.
We will refer to $W_{L,R}$ and $M_{L,R}$ as the (nonhermitian) topological invariant and the (nonhermitian) topological mass on opposite sides of the smooth domain wall.
These quantities depend only on the asymtpotic values of the scalar fields ${m}_{L,R}$ and ${v}_{L,R}$.
For the asymptotically hermitian case $\Im({m}_{L,R})=\Im({v}_{L,R})=0$, one has that $\sgn({M}_{L,R})=\sgn({m}_{L,R})$, ${K}_{L,R}=|\Im({q}_{L,R})|$, and ${W}_{L,R}=\sgn({v}_{L,R})$ for ${m}_{L,R}\le0$ and ${W}_{L,R}=0$ otherwise.
Hence, our definitions of nonhermitian topological mass $M_{L,R}$ and nonhermitian topological invariant $W_{L,R}$ in \cref{eq:determinants} reduces to the usual definition of topological mass and topological invariant in the hermitian case.

To demonstrate the relations between the (nonhermitian) topological invariants $W_{L,R}$ and the existence and number of zero modes, we start to analyze the asymptotic behavior of the solutions of the Jackiw-Rebbi equation.
Since the two solutions decay as $|\varphi_{1,2}^s(x\to-\infty)|={e}^{\Re(\alpha_L^\pm) x}$ and $\varphi_{1,2}^s(x\to+\infty)={e}^{-\Re(\alpha_R^\mp) x}$, the boundary conditions on the left are satisfied when $\Re(\alpha_L^\pm)>0$
while the boundary conditions on the right are satisfied when $\Re(\alpha_R^\mp)>0$, respectively for the two solutions $\varphi_{1,2}^s(x)$.
To determine whether the solutions satisfy the boundary conditions, we thus need to study the sign of the exponents $\Re(\alpha_{L,R}^\pm)$.
For ${m}_L,{v}_L\in\mathbb{C}$, these exponents are completely determined by the quantities ${M}_L$ and ${K}_L$, since
\begin{equation}
\sgn(\Re(\alpha_L^\pm))=
\left\{
\begin{array}{ll}
\pm1, 
&\text{ for } {M}_L>0,
\\
0 \text{ or } -s W_L, 
&\text{ for } {M}_L=0,
\\
-s W_L,
&\text{ for } {M}_L<0.
\end{array}
\right.
\label{eq:alphamucondition}
\end{equation}
In the first case, the solution $\varphi_1^s(x)$ decays while the solution $\varphi_2^s(x)$ diverges for $x\to-\infty$. 
In the second case, one of the solutions $\varphi_{1,2}^s(x)$ converges to a constant value.
In the last case, both solutions $\varphi_{1,2}^s(x)$ decays for $x\to-\infty$ as long as $s=-\sgn(\Re({v}_L))$.
Analogous statements hold for $\alpha_R^\pm$ by exchanging left and right ${v}_L\to{v}_R$, ${m}_L\to{m}_R$, $W_L\to W_R$, ${M}_L\to {M}_R$, and $s\to-s$, that is
\begin{equation}
\sgn(\Re(\alpha_R^\pm))=
\left\{
\begin{array}{ll}
\pm1, 
&\text{ for } {M}_R>0,
\\
0 \text{ or } -s W_R, 
&\text{ for } {M}_R=0,
\\
s W_R,
&\text{ for } {M}_R<0.
\end{array}
\right.
\label{eq:alphamucondition2}
\end{equation}

We are now ready to demonstrate the nonhermitian generalization of the bulk-boundary correspondence.
To do so, we will demonstrate that the number of localized zero-energy modes in an infinite-size system defined on the interval $(-\infty,\infty)$ is equal to $|W_L-W_R|$, and that the number of localized zero-energy modes in a semi-infinite system defined on the interval $[0,\infty)$ is equal to $|W_R|$.

\subsection{Infinite interval $(-\infty,\infty)$  \label{sec:JR-infinf}}

We now consider zero-energy modes in an infinite-size system.
These are defined as the eigenmodes with zero energy that satisfy the boundary conditions $\varphi(\pm\infty)=0$ on the interval $(-\infty,\infty)$.
In order to satisfy the boundary conditions, the modes need to decay asymptotically for $x\to\pm\infty$, which mandates that the exponents need to be $\Re(\alpha_L^\pm)>0$ and $\Re(\alpha_R^\pm)>0$.
In this case, there exists $|W_L-W_R|$ zero modes, i.e., two modes if $W_{L}W_R=-1$, and one mode if $|W_{L}|=1$ and $W_R=0$ or $|W_{R}|=1$ and $W_L=0$.
To demonstrate these statements, we need to find the particular solutions that satisfy 
$\varphi(x\to\pm\infty)=0$, i.e., $\varphi(y\to0)=\varphi(y\to1)=0$.
Assuming $|F_{1,2}(y)|$ convergent or logarithmically divergent at $y=0,1$, we need to care about the factors $y^{w\alpha_L^-}$ and $(1-y)^{w\alpha_R^-}$, which converge to zero at $y=0$ and $y=1$ only if $\Re({\alpha_L^-})>0$ and $\Re({\alpha_R^-})>0$, respectively.
This gives only four possible outcomes, discussed below and summarized in \cref{tab:solintervals}.

\subsubsection{Two zero modes for $W_LW_R=-1$\label{sec:twosolutions}}

For $\sgn({M}_L)=\sgn({M}_R)=-1$, and $\Re({v}_L)\Re({v}_R)<0$, the topological invariants are opposite and nonzero $W_LW_R=-1$.
Moreover, in this case, one has
$\sgn(\Re(\alpha_L^\pm))=-s\sgn(\Re({v}_L))$ and
$\sgn(\Re(\alpha_R^\pm))=s\sgn(\Re({v}_R))$,
which gives 
$\sgn(\Re(\alpha_L^\pm))=\sgn(\Re(\alpha_R^\pm))$ in this case.
Choosing the pseudospin $s=\sgn{(\Re({v}_L))}=-\sgn{\Re(({v}_R))}$ gives $\Re(s{v}_L)>0$ and $\Re(s{v}_R)<0$ and therefore $\Re(\alpha_L^\pm)$ and $\Re(\alpha_R^\pm)<0$:
In this case, the two linearly independent solutions $\varphi_{1,2}^s(x)$ in \cref{eq:2generalsolutions} diverge exponentially with different decaying lengths at $x\to\pm\infty$, and their linear combination cannot satisfy the boundary conditions.
Choosing instead the opposite pseudospin $s=-\sgn{(\Re({v}_L))}=\sgn{(\Re({v}_R))}$ gives $\Re(s{v}_L)<0$ and $\Re(s{v}_R)>0$ and therefore $\Re(\alpha_L^\pm)$ and $\Re(\alpha_R^\pm)>0$:
In this case, the two solutions $\varphi_{1,2}^s(x)$ decay exponentially for $x\to\pm\infty$.
Therefore, the eigenmodes satisfying the boundary conditions are
\begin{equation}
	\varphi(x)=A_1\varphi_1^s(x)+A_2\varphi_2^s(x),
\end{equation}
up to normalization constants $A_{1,2}$, with pseudospin $s=-\sgn{(\Re({v}_L))}=\sgn{(\Re({v}_R))}$ and $\varphi_{1,2}^s(x)$ as in \cref{eq:2generalsolutions,eq:2generalsolutionshyper}, and ${\alpha_L}$, ${\alpha_R}$ given by \cref{eq:alphabeta}.
Since $\Re(\alpha_{L,R}^-)\le\Re(\alpha_{L,R}^+)$, the asymptotic behavior is 
\begin{align}
\varphi(x\to-\infty)&\sim {e}^{+\alpha_L^- x}={e}^{+\mu_L^- x}{e}^{+\mathrm{i}\kappa_L^- x},
\\
\varphi(x\to+\infty)&\sim {e}^{-\alpha_R^- x}={e}^{-\mu_R^- x}{e}^{-\mathrm{i}\kappa_R^- x},
\end{align}
where $\mu_{L,R}^-=\Re(\alpha_{L,R}^-)=-{M}_{L,R}$.

The asymptotic behavior of the two modes $\varphi_1^s$ and $\varphi_2^s$ on the left of the smooth domain wall is determined by the exponents 
$\alpha_{L}^+=-s{v}_{L}+{q}_{L}$ (decay length $\xi_L^+$ and oscillation wavelength $\lambda_L^+$) and 
$\alpha_{L}^-=-s{v}_{L}-{q}_{L}$ (decay length $\xi_L^-$ and oscillation wavelength $\lambda_L^-$), respectively,
and on the right by the exponents 
$\alpha_{R}^-=s{v}_{R}-{q}_{R}$ (decay length $\xi_R^-$ and oscillation wavelength $\lambda_R^-$) and
$\alpha_{R}^+=s{v}_{R}+{q}_{R}$ (decay length $\xi_R^+$ and oscillation wavelength $\lambda_R^+$), respectively.

\subsubsection{One zero mode for $W_L=0$ and $|W_R|=1$}

For $\sgn({M}_L)=1$, $\sgn({M}_R)=-1$, and $\Re({v}_R)\neq0$, the topological invariants are $W_L=0$ and $W_R=\sgn(\Re({v}_R))=\pm1$.
Moreover, in this case, one has $\sgn(\Re(\alpha_R^\pm))=s\sgn(\Re({v}_R))$ and $\Re(\alpha_L^+)>0>\Re(\alpha_L^-)$.
Choosing $s=-\sgn(\Re({v}_R))$ gives $\Re(\alpha_R^\pm)<0$, and therefore the two linearly independent solutions $\varphi_{1,2}^s(x)$ in \cref{eq:2generalsolutions} respectively decay and diverge for $x\to-\infty$ and both diverge for $x\to\infty$, and thus their linear combination cannot satisfy the boundary conditions.
Choosing instead $s=\sgn(\Re({v}_R))$ gives $\Re(\alpha_R^\pm)>0$, and consequently the solution $\varphi_1^s(x)$ now decays for $x\to\pm\infty$, while $\varphi_2^s(x)$ still diverges for $x\to-\infty$.
Hence, the eigenmode satisfying the boundary conditions is 
\begin{equation} \label{eq:1solphi1}
	\varphi(x)=
	\varphi_1^s(x)
	,
\end{equation}
up to a normalization constant, with pseudospin $s=\sgn(\Re({v}_R))$ and $\varphi_1^s(x)$ as in \cref{eq:2generalsolutions,eq:2generalsolutionshyper}, and ${\alpha_L}$, ${\alpha_R}$ given by \cref{eq:alphabeta}.
The asymptotic behavior is
\begin{align}
\varphi(x\to-\infty)&\sim {e}^{+\alpha_L^+ x}={e}^{+\mu_L^+ x}{e}^{+\mathrm{i}\kappa_L^+ x},
\\
\varphi(x\to+\infty)&\sim {e}^{-\alpha_R^- x}={e}^{-\mu_R^- x}{e}^{-\mathrm{i}\kappa_R^- x},
\end{align}
where $\mu_{R}^-=\Re(\alpha_{R}^-)=-{M}_{R}$.

For the limiting case $\sgn({M}_L)= 0$, $\sgn({M}_R)=-1$, and $\Re({v}_L)\Re({v}_R)<0$, the solution in \cref{eq:1solphi1} still satisfy the boundary conditions.
Indeed, by taking ${m}_L= 0$ into the solution, one has that $\alpha_L^+>0$ if $s=-\sgn(\Re({v}_L))$, which is indeed the case since we choose $s=\sgn(\Re({v}_R))$ and $\Re({v}_L)\Re({v}_R)<0$.

The asymptotic behavior on the left is determined by the exponents $\alpha_{L}^+=-s{v}_{L}+{q}_{L}$ (decay length $\xi_L^+$ and oscillation wavelength $\lambda_L^+$) and on the right by $\alpha_{R}^-=s{v}_{R}-{q}_{R}$ (decay length $\xi_R^-$ and oscillation wavelength $\lambda_R^-$).

\subsubsection{One zero mode for $|W_L|=1$, $W_R=0$}

Analogous arguments hold if
$\sgn({M}_L)=-1$, $\sgn({M}_R)=1$, and $\Re({v}_L)\neq0$, where the topological invariants are $W_L=\sgn(\Re({v}_L))=\pm1$ and $W_R=0$.
Moreover, in this case, one has $\sgn(\Re(\alpha_L^\pm))=-s\sgn(\Re({v}_L))$ and $\Re(\alpha_R^+)>0>\Re(\alpha_R^-)$.
Choosing $s=\sgn(\Re({v}_L))$ yields $\Re(\alpha_L^\pm)<0$, and therefore the two linearly independent solutions $\varphi_{1,2}^s(x)$ in \cref{eq:2generalsolutions} respectively diverge and decays for $x\to\infty$ and both diverge for $x\to-\infty$, and thus their linear combination cannot satisfy the boundary conditions.
Choosing instead $s=-\sgn(\Re({v}_L))$ gives $\Re(\alpha_L^\pm)>0$, and therefore the solution $\varphi_2^s(x)$ now decays for $x\to\pm\infty$ while $\varphi_1^s(x)$ diverges for $x\to\infty$.
Hence, the eigenmode satisfying the boundary conditions is 
\begin{equation} \label{eq:1solphi2}
	\varphi(x)=
	\varphi_2^s(x)
\end{equation}
up to a normalization constant, with corresponding pseudospin $s=-\sgn(\Re({v}_L))$ and $\varphi_2^s(x)$ as in \cref{eq:2generalsolutions,eq:2generalsolutionshyper}, and ${\alpha_L}$, ${\alpha_R}$ given by \cref{eq:alphabeta}.
The asymptotic behavior is
\begin{align}
\varphi(x\to-\infty)&\sim {e}^{+\alpha_L^- x}={e}^{+\mu_L^- x}{e}^{+\mathrm{i}\kappa_L^- x},
\\
\varphi(x\to+\infty)&\sim {e}^{-\alpha_R^+ x}={e}^{-\mu_R^+ x}{e}^{-\mathrm{i}\kappa_R^+ x},
\end{align}
where $\mu_{L}^-=\Re(\alpha_{L}^-)=-{M}_{L}$.
It is clear that this case is equivalent to the previous one up to the transformation $x\to-x$, (i.e., $y\to1-y$), which corresponds also to substituting $s\to-s$, ${\alpha_L}\leftrightarrow{\alpha_R}$, ${v}_L\leftrightarrow{v}_R$, and ${m}_L\leftrightarrow{m}_R$.
This gives also ${v}_1\to-{v}_1$ and ${m}_2\to{m}_2$ in \cref{eq:expansion} which yields $a,b\to a,b$ and $1-c\to c-a-b$.

For the limiting case $\sgn({M}_L)=-1$, $\sgn({M}_R)=0$, and $\Re({v}_L)\Re({v}_R)<0$, the solution in \cref{eq:1solphi2} still satisfy the boundary conditions, by taking ${m}_R= 0$ into the solution, giving ${K}_R^+>0$ if $s=\sgn(\Re({v}_R))$, which is indeed the case since we choose $s=-\sgn(\Re({v}_L))$ and $\Re({v}_L)\Re({v}_R)<0$.

The asymptotic behavior on the left is determined by the exponents $\alpha_{L}^-=-s{v}_{L}-{q}_{L}$ (decay length $\xi_L^-$ and oscillation wavelength $\lambda_L^-$) and on the right by $\alpha_{R}^+=s{v}_{R}+{q}_{R}$ (decay length $\xi_R^+$ and oscillation wavelength $\lambda_R^+$).

\subsubsection{No zero modes otherwise}

In any other case, no linear combination of the general solutions satisfies the boundary conditions with the exception of the trivial solution $\varphi(x)=0$.

For $\sgn({M}_L)\ge0$ and $\sgn({M}_R)\ge0$, the topological invariants are zero $W_{L,R}=0$, and one has $\Re(\alpha_L^+)\ge0\ge\Re(\alpha_L^-)$ and $\Re(\alpha_R^+)\ge0\ge\Re(\alpha_R^-)$.
The solution $\varphi_1^s(x)$ decays only for $x\to-\infty$, while the solution $\varphi_2^s(y)$ decays only for $x\to\infty$.
Therefore, one cannot find any nontrivial linear combination that satisfies the boundary conditions.

For $\sgn({M}_L)=-1$, $\sgn({M}_R)=-1$, and $\Re({v}_L)\Re({v}_R)>0$, the topological invariants are equal and nonzero $W_L=W_R=\pm1$, and one has $\sgn(\Re(\alpha_L^\pm))=-s\sgn(\Re({v}_L))$ and $\sgn(\Re(\alpha_R^\pm))=s\sgn(\Re({v}_R))$, which gives $\sgn(\Re(\alpha_L^\pm))=-\sgn(\Re(\alpha_R^\pm))$ in this case.
Choosing the pseudospin such that $\Re(\alpha_R^\pm)>0$, one has that $\varphi_{1,2}^s(x)$ diverge for $x\to-\infty$ with different decaying lengths, so that no nontrivial linear combination can satisfy the boundary conditions.
Conversely, choosing the pseudospin such that $\Re(\alpha_L^\pm)>0$ one has that $\varphi_{1,2}^s(x)$ diverge for $x\to\infty$ with different decaying lengths, so that no nontrivial linear combination can satisfy the boundary conditions.
Hence, the boundary conditions cannot be satisfied.

In the case where 
$\sgn({M}_L)=0$, $\sgn({M}_R)\le0$, and $\Re({v}_L)\Re({v}_R)\ge0$, 
the case
$\sgn({M}_L)\le0$, $\sgn({M}_R)=0$, and $\Re({v}_L)\Re({v}_R)\ge0$, 
and the case
$\sgn({M}_L)\le0$, $\sgn({M}_R)\le0$, and $\Re({v}_L)\Re({v}_R)=0$,
at least one of the four exponents $\alpha_{L,R}^\pm$ vanishes so that the corresponding asymptote converges to a constant value.
Therefore, $\varphi_{1,2}^s(x)$ diverge for $x\to\pm\infty$ with different decaying lengths or converge to a constant, and again no nontrivial linear combination can satisfy the boundary conditions.
Hence, the boundary conditions cannot be satisfied for ${m}_L\le0$, ${m}_R\le0$, and $\Re({v}_L)\Re({v}_R)\ge0$.

Moreover, in the case where $\sgn({M}_L)=1$, $\sgn({M}_R)=-1$, and $\Re({v}_R)\Re({v}_L)=0$, the solution \cref{eq:1solphi1} does not decay for either $x\to\pm\infty$.
Finally, in the case where $\sgn({M}_L)=-1$, $\sgn({M}_R)=1$, and $\Re({v}_R)\Re({v}_L)=0$, the solution \cref{eq:1solphi2} does not decay for either $x\to\pm\infty$, and also in these cases, the boundary conditions cannot be satisfied.

\subsection{Semi-infinite interval $[0,\infty)$ \label{sec:JR-0inf}}

We now consider zero-energy modes in a semi-infinite system.
These are defined as the eigenmodes with zero energy that satisfy the boundary conditions $\varphi(0)=\varphi(\infty)=0$ on the interval $[0,\infty)$.
In order to satisfy the boundary conditions, the modes need to decay asymptotically for $x\to\infty$, which mandates that $\Re(\alpha_R^\pm)>0$.
As expected, there is one independent mode if $W_R\neq1$.
To demonstrate this statement, we need to find the particular solutions that satisfy the boundary conditions $\varphi(x\to0)=\varphi(x\to\infty)=0$, i.e., $\varphi(y\to1/2)=\varphi(y\to1)=0$.

Assuming  $|F_{1,2}(y)|$ convergent or logarithmically divergent at $y=1/2,1$, we only need to care now about the factor $(1-y)^{w\alpha_R^-}$, which converge to zero at $y=1$ only if $\Re({\alpha_R^-})>0$.
This gives only two possible outcomes, discussed below and summarized in \cref{tab:solintervals}.
Zero modes on the interval $(-\infty,0]$ are obtained similarly.
Note that the confinement to semi-infinite intervals introduces a sharp domain wall at $x=0$, being effectively equivalent to an infinite potential wall superimposed to a smooth domain wall with a finite width when $w>0$.

\subsubsection{One zero mode for $W_R=\pm1$}

For $\sgn({M}_R)=-1$ and $\Re({v}_R)\neq0$, the topological invariant is nonzero $W_R=\sgn(\Re({v}_R))\neq0$, giving $\sgn(\Re(\alpha_R^\pm))=s\sgn(\Re({v}_R))$.
Choosing $s=-\sgn(\Re({v}_R))$ gives $\Re(\alpha_R^\pm)<0$, and therefore the two linearly independent solutions $\varphi_{1,2}^s(x)$ in \cref{eq:2generalsolutions} diverge for $x\to\infty$ with different decaying lengths, and thus their linear combination cannot satisfy the boundary conditions.
Choosing instead $s=\sgn(\Re({v}_R))$ gives $\Re(\alpha_R^\pm)>0$, and with both solutions $\varphi_{1,2}^s(x)$ decaying to zero for $x\to\infty$.
Therefore, the eigenmode satisfying the boundary conditions is 
\begin{equation}\label{eq:semiinfinitesolution}
	\varphi(x)=A_1\varphi_1^s(x)+A_2\varphi_2^s(x),
\end{equation}
with pseudospin $s=\sgn(\Re({v}_R))$ and $\varphi_{1,2}^s(x)$ as in \cref{eq:2generalsolutions,eq:2generalsolutionshyper}, ${\alpha_L}$, ${\alpha_R}$ given by \cref{eq:alphabeta}, and with $A_{1,2}$ chosen such that $\varphi(0)=A_1\varphi_1^s(0)+A_2\varphi_2^s(0)=0$.
Since $\Re(\alpha_{R}^-)\le\Re(\alpha_{R}^+)$, the asymptotic behavior is
\begin{align}
\varphi(x\to+\infty)&\sim {e}^{-\alpha_R^- x}={e}^{-\mu_R^- x}{e}^{-\mathrm{i}\kappa_R^- x},
\end{align}
where $\mu_{R}^-=\Re(\alpha_{R}^-)=-{M}_{R}$.
If one can expand ${m}(x)$ and ${v}(x)$ as in \cref{eq:expansion}, taking $F_{1,2}$ as in \cref{eq:2generalsolutionshyper}, the constants $A_{1,2}$ must satisfy
\begin{equation}\label{ew:A12}
A_1\frac{{{\,}_2F_1}\big(a_1,b_1,c_1,1/2\big) }{2^{{w\alpha_L^+}+{w\alpha_R^-}} }
+
A_2\frac{{{\,}_2F_1}\big(a_2,b_2,c_2,1/2\big) }{2^{{w\alpha_L^-}+{w\alpha_R^+}} }
=0.
\end{equation}

Notice that, when considering the eigenmodes of the modified Jackiw-Rebbi equation on the interval $[0,\infty)$, the choice of the fields at $x=-\infty$ is somewhat arbitrary.
Therefore, one can always choose the values of ${v}_L$ and ${m}_L$ in order to match the conditions 
${v}_{L,R}={v}$,
${v}_{L,R}={v}$ and ${m}_{L,R}={m}$,
${v}_{L,R}=\pm{v}$,
${v}_{L,R}=\pm{v}$ and ${m}_{L,R}={m}$ in Table I of Ref.~\cite{marra_topological_2024}.
and obtain the respective simplified forms of the wavefunctions.
However, if one can expand ${m}(x)$ and ${v}(x)$ as in \cref{eq:expansion}, the functional form of the fields ${m}(x)$ and ${v}(x)$ on the interval $[0,\infty)$ determines univocally the values of the fields at $x=-\infty$.

The asymptotic behavior is determined by the exponent $\alpha_{R}^-=s{v}_{R}-{q}_{R}$ (decay length $\xi_R^-$ and oscillation wavelength $\lambda_R^-$).
This is because, even though the two solutions have different decaying lengths $\xi_R^\pm$, one has that $\xi_R^-\ge \xi_R^+$, and therefore the decaying length $\xi_R^-$ dominates at large distances.
In the hermitian case $\Im({m}(x))=\Im({v}(x))=0$, it is always possible to find a set of $A_{1,2}$ satisfying the equations above and such that $\varphi(x)$ is real.

\subsubsection{No zero modes otherwise}

If $\sgn({M}_R)\ge0$ or if $\Re({v}_R)=0$, no nontrivial linear combination of the general solutions can satisfy the boundary conditions.
For $\sgn({M}_R)=1$, the topological invariant is zero $W_R=0$, and one has $\Re(\alpha_R^+)>0>\Re(\alpha_R^-)$.
Therefore only the solution $\varphi_2^s(x)$ in \cref{eq:2generalsolutions} decays for $x\to\infty$.
For $\sgn({M}_R)=0$, one has that either $\alpha_R^+=0$ or $\alpha_R$, depending on the choice of the pseudospin.
Therefore, only one of the solutions $\varphi_{1,2}^s(x)$ decays for $x\to\infty$.
However, since in general $\varphi_{1,2}^s(0)\neq0$, the boundary conditions cannot be satisfied.
For ${m}_R<0$ (or $\sgn({M}_R)=-1$) and $\Re({v}_R)=0$, one has $\Re(\alpha_R^\pm)=0$.
Hence, neither solutions $\varphi_{1,2}^s(x)$ decay for $x\to\infty$ and thus the boundary conditions cannot be satisfied.

\section{Properties of the zero modes}

\subsection{Oscillating behavior}

Here, we will see that the quantity ${K}_{L,R}$ will discriminate between oscillatory and nonoscillatory behavior, generalizing to the nonhermitian case the analogous quantity introduced in Ref.~\cite{marra_topological_2024} for hermitian systems.
One has
\begin{equation}
\begin{array}{ll}
\Im(\alpha_L^\pm)=0, 
&\text{ for } {K}_L=0,
\\
\Im(\alpha_L^\pm)\neq0
&\text{ for } {K}_L\neq0,
\end{array}
\end{equation}
where we remind that $\Im(\alpha_L^\pm)=\kappa_L^\pm$.
One has also
\begin{equation}
\begin{array}{ll}
\Re(\alpha_L^+)=\Re(\alpha_L^-), &\text{ if } \Re({q}_L)=0,
\\
\Im(\alpha_L^+)=-\Im(\alpha_L^-), &\text{ if } \Im({v}_L)=0,
\end{array}
\end{equation}
where we remind that $\Re(\alpha_L^\pm)=\mu_L^\pm$ and $\Im(\alpha_L^\pm)=\kappa_L^\pm$.
The case where ${K}_L\neq0$ corresponds to oscillating behavior for $x\to-\infty$, while ${K}_L=0$ corresponds to no oscillations.
If ${m}_L\in\mathbb{R}$, we see that $\sgn({M}_L)=\sgn({m}_L)$, and thus the conditions in \cref{eq:alphamucondition} simplify by substituting ${M}_L\to{m}_L$
(analogously the conditions in \cref{eq:alphamucondition2} simplify by substituting ${M}_R\to{m}_R$).
Moreover, if the system is asymptotically hermitian for $x\to-\infty$, i.e., if ${m}_L,{v}_L\in\mathbb{R}$, one has that ${K}_L=0$ if ${{v}}_L^2+{m}_L\ge0$ and ${K}_L\neq0$ otherwise.
This gives 
\begin{equation}
\begin{array}{ll}
\Im(\alpha_L^\pm)=0, 
&\text{ for } {{v}}_L^2+{m}_L>0,
\\
\Im(\alpha_L^\pm)=0, \quad\Re(\alpha_L^+)=\Re(\alpha_L^-), 
&\text{ for } {{v}}_L^2+{m}_L=0,
\\
\Im(\alpha_L^\pm)\neq0,
\quad
\Im(\alpha_L^+)=-\Im(\alpha_L^-), 
&\text{ for } {{v}}_L^2+{m}_L<0,
\end{array}
\end{equation}
for ${m}_L,{v}_L\in\mathbb{R}$.
Analogous statements hold for $\alpha_R^\pm$ by exchanging left and right ${v}_L\to{v}_R$, and ${m}_L\to{m}_R$.

The oscillatory behavior of the eigenmodes is determined by ${K}_{L,R}$.
For ${K}_{L,R}=0$, one gets $\kappa_{L,R}=0$, respectively, which mandates that the modes decay exponentially for $x\to\mp\infty$, respectively.
Conversely, for ${K}_{L,R}\neq0$, one has $\kappa_{L,R}\neq0$, respectively, which mandates modes with exponentially damped oscillations for $x\to\mp\infty$, respectively.
In short, one has
\begin{equation}
\begin{cases}
{K}_{L,R}=0 \rightarrow \text{exp.~decay}
\\
{K}_{L,R}\neq0 \rightarrow \text{exp.~damped oscill.}
\end{cases}
\text{ for } x\to\mp\infty
\end{equation}
This is also summarized in \cref{tab:solintervals}.

\subsection{Reality}

Note also that, in the hermitian case $\Im({m}(x))=\Im({v}(x))=0$, the modified Jackiw-Rebbi equation is real, and consequently, all eigenmodes are real for some choices of $A_{1,2}$.
In the nonhermitian case when $\Im({m}(x))\neq0$ or $\Im({v}(x))\neq0$ instead, it is not possible to choose $A_{1,2}$ such that the solutions are real:
In other words, the solutions are complex for all choices of $A_{1,2}$.
This also mandates that the complex phase of the wavefunction is not uniform in the nonhermitian case.

\subsection{Length scales and hairstyles}

As we have shown, the existence and number of zero modes only depend on the topological invariants $W_{L,R}$ and ultimately on the asymptotic values of the fields ${m}_{L,R}$ and ${v}_{L,R}$.
These results generalize the bulk-boundary correspondence~\cite{hatsugai_chern_1993,ryu_topological_2002,teo_topological_2010} to nonhermitian systems.
On the other hand, the quantities $W_{L,R}$ in \cref{eq:TI} generalize the topological invariants to nonhermitian systems.
Moreover, the asymptotic behavior of each mode in \cref{eq:decayMAIN} is determined only by the decay rates $\mu_{L,R}^\pm$ and oscillation momenta $\kappa_{L,R}^\pm$ (or, alternatively, by the decay lengths $\xi_{L,R}^\pm$ and oscillation wavelengths $\lambda_{L,R}^\pm$), which ultimately depend only on the asymptotic values of the fields ${m}_{L,R}$ and ${v}_{L,R}$.
On the left asymptote, the case ${K}_L=0$ mandates $\kappa_{L}^\pm=0$ (i.e., $\lambda_{L}^\pm=\infty$), which corresponds to an exponential decay (without oscillations), while the case ${K}_L\neq0$ mandates $\kappa_{L}^\pm\neq0$ (i.e., $\lambda_{L}^\pm<\infty$) which correspond to exponentially damped oscillations.
If $\Im({m}_{L})=\Im({v}_{L})=0$, the case ${K}_L=0$ is realized for ${v}_L^2+{m}_L\ge0$ (exponential decay) while the case ${K}_L\neq0$ is realized for ${v}_L^2+{m}_L<0$ (exponentially damped oscillations).
Moreover, one has that $\mu_L^+=\mu_L^-$ if $\Re({q}_L)=0$, and $\kappa_L^+=-\kappa_L^-$ if $\Im({v}_L)=0$.
Analogous statements hold for the right asymptote.

\begin{table*}[t]
 \centering
 \begin{tabular}{@{} llll @{}}
\multicolumn{4}{c}{infinite interval $(-\infty,\infty)$} \\\\[-2mm]
\hline\\
$W_LW_R=-1$ &\qquad\qquad& 
\multirow{2}{*}{2 modes 
$\begin{cases}
\varphi_{1}^s(x), 
\\
\varphi_{2}^s(x), 
\end{cases}$}
& 
\multirow{2}{*}{$s=-W_L=W_R$}
\\
${M}_L{M}_R\neq0$
\\\\\hline\\
$W_L=0$, $|W_R|=1$ \\
${M}_L{M}_R\neq0$
\\
&& 1 mode $\varphi_{1}^s(x)$ 
&
$s=W_R$
\\
$W_LW_R=-1$ (limiting case) \\
${M}_L=0$, ${M}_R\neq0$
\\
\\\hline\\
$|W_L|=1$, $W_R=0$ \\
${M}_L{M}_R\neq0$
\\ 
&& 1 mode $\varphi_{2}^s(x)$ 
& 
$s=-W_L$
\\
$W_LW_R=-1$ (limiting case) \\
${M}_L\neq0$, ${M}_R=0$
\\
\\\hline\\
otherwise && none \\
\\\hline\hline\\\\
\multicolumn{4}{c}{semi-infinite interval $[0,\infty)$} \\\\[-2mm]
\hline\\
$|W_R|=1$ && 
\multirow{2}{*}{1 mode $A_1\varphi_1^s(x)+A_2\varphi_2^s(x)$ 
\quad 
} 
& 
\multirow{2}{*}{$s=W_R$}\\
${M}_R\neq0$
\\
\\\hline\\
otherwise && none \\
\\\hline\hline\\
 \end{tabular}
\caption{
The number of eigenmodes $\varphi_{1,2}^s(x)$ satisfying the boundary conditions and their corresponding pseudospin $s$ on the infinite interval $(-\infty,\infty)$ and on the semi-infinite interval $[0,\infty)$ for different regimes determined by the quantities ${M}_{L,R}$, ${K}_{L,R}$, and $W_{L,R}$ (topological invariant) defined in \cref{eq:determinants}, which ultimately depend on the the asymptotic values of the fields ${m}_{L,R},{v}_{L,R}\in\mathbb{C}$.
The modes have exponential decay when ${K}_{L,R}=0$ or exponentially damped oscillations when ${K}_{L,R}\neq0$ for $x\to\mp\infty$.
In the asymtpotical hermitian regime (i.e., hermitian at large distances) where ${m}_{L,R},{v}_{L,R}\in\mathbb{R}$, which gives $\sgn({M}_{L,R})=\sgn({m}_{L,R})$ and ${K}_{L,R}=|\Im({q}_{L,R})|$, the modes have exponential decay when ${{v}}_L^2+{m}_L\ge0$ or exponentially damped oscillations when ${{v}}_L^2+{m}_L<0$.
The asymptotic behavior is fully determined by the decay rates $\mu_{L,R}^\pm=\Re(\alpha_{L,R}^\pm)$ and the oscillation momenta $\kappa_{L,R}^\pm=\Im(\alpha_{L,R}^\pm)$, or alternatively by the decay lengths $\xi_{L,R}^\pm=1/\Re(\alpha_{L,R}^\pm)$ and oscillation wavelengths $\lambda_{L,R}^\pm=2\pi/\Im(\alpha_{L,R}^\pm)$.
Moreover, $\mu_{L,R}^+=\mu_{L,R}^-=\mp s\Re({v}_{L,R})$ if $\Re({q}_{L,R})=0$ and $\kappa_{L,R}^+=-\kappa_{L,R}^-=\Im({q}_{L,R})$ if $\Im({v}_{L,R})=0$.
In the asymptotically hermitian regime, one has
$\kappa_{L,R}^+=\kappa_{L,R}^-=0$ for ${{v}}_{L,R}^2+{m}_{L,R}\ge0$,
with
$\mu_{L,R}^+=\mu_{L,R}^-$ for ${{v}}_{L,R}^2+{m}_{L,R}=0$,
while
$\mu_{L,R}^+=\mu_{L,R}^-=\mp s{v}_{L,R}$ and
$\kappa_{L,R}^+=-\kappa_{L,R}^-=\Im({q}_{L,R})\neq0$ for ${{v}}_{L,R}^2+{m}_{L,R}<0$.
This Table summarizes the existence and number of zero-energy eigenmodes of the Jackiw-Rebbi equation in the nonhermitian case.
The hermitian case can be obtained as a special case.
For a direct comparison to the hermitian case, see Table II of Ref.~\cite{marra_zero_2025}.
}
 \label{tab:solintervals}
\end{table*}

In the case of a sharp domain wall $w\to0$, the zero modes are featureless objects with "no hair" fully described by $\mu_{L,R}^\pm$ and  $\kappa_{L,R}^\pm$ (or, alternatively, by $\xi_{L,R}^\pm$ and   $\lambda_{L,R}^\pm$), to use the terminology introduced in Ref.~\cite{marra_topological_2024}.
In the case of a smooth domain wall, however, the zero modes are nonfeatureless, since they cannot be fully described by $\mu_{L,R}^\pm$ and  $\kappa_{L,R}^\pm$ (or by $\xi_{L,R}^\pm$ and $\lambda_{L,R}^\pm$), especially in the region close to the domain wall $|x|<w$. 
We further distinguish the cases of zero modes with "short hair" when $w<\xi$ and with "long hair" when $w>\xi$.
Zero modes with short hair appear featureless at large distances $|x|>\xi>w$ but not at small distances $|x|<w$, where their wavefunctions crucially depend on the detail of the fields near the domain wall.
Zero modes with long hair have instead wavefunctions that are nonfeatureless at all length scales and crucially depend on the details of the fields.
Notice also that the property of having short or long hair depends on the side considered since a zero mode can have short hair on the left $x<0$ and long hair on the right of the domain wall $x>0$, or vice versa.
The existence of zero modes localized at smooth domain walls is still a direct consequence of the bulk-boundary correspondence in the hermitian regime and of the generalized form of the bulk-boundary correspondence (in the language of the generalized invariant in \cref{eq:TI}) in the nonhermitian regime.

We stress that the terminology used here is meant to be taken with a grain of salt.
Black holes are said to have no hair since each black hole can be completely described (in the frame of reference of its center of mass) by only a few physical quantities, namely, its mass, electric charge, and angular momentum.
Similarly, zero modes with no hair can be completely described by only a few physical quantities, namely the decay rates $\mu_{L,R}^\pm$ and oscillation momenta $\kappa_{L,R}^\pm$ (or alternatively decay lengths $\xi_{L,R}^\pm$ and oscillation wavelengths $\lambda_{L,R}^\pm$).
Zero modes with hair cannot be fully described by these quantities alone.
However, there is a fundamental difference between hairless zero modes and black holes:
Zero modes are local objects, while black holes are not.
Indeed, the wavefunctions of zero modes decay exponentially, and therefore zero modes cannot be directly observable at a distance much larger than their localization lengths.
It is simply not possible to measure their decay rates and oscillation momenta at a distance larger than their localization length.
In contrast, a black hole induces gravitational and electromagnetic fields which scale polynomially and not exponentially, and consequently its mass, charge, and angular momentum are observable at all distances and independent from the observer's distance.

\subsection{Local properties and local topological invariant}

In both featureless and nonfeatureless cases, the long-distance behavior $|x|\gg w$ of the zero modes is completely described by \cref{eq:decayMAIN}.
In particular, the different asymptotic regimes are completely determined by the quantities ${M}_{L,R}$, ${K}_{L,R}$, and $W_{L,R}$ defined in \cref{eq:determinants} which are in turn determined by the asymtpotic values of the fields ${m}(x)$ and ${v}(x)$ at long distances $x\to\pm\infty$. 
However, to understand the properties of the wavefunction at short distances $|x|\lesssim w$, we introduce the following functions
\begin{subequations}\label{eq:determinantsRS}
\begin{align}
{M}(x)=&
|\Re(\sqrt{{v}(x)^2+{m}(x)})|-|\Re({v}(x))|,
\\
{K}(x)=&
|\Im(\sqrt{{v}(x)^2+{m}(x)})|+|\Im({v}(x))|,
\\\label{eq:TIRS}
W(x)=&
\begin{cases}
0 & \text{ if } \sgn({M}(x))=1.
\\
\sgn(\Re({v}(x))), & \text{ otherwise},
\end{cases}
\end{align}
\end{subequations}
which generalize the quantities ${M}_{L,R}$, ${K}_{L,R}$, and $W_{L,R}$. 
We will refer to $W(x)$ as the \emph{local} topological invariant and to $M(x)$ as the \emph{local} topological mass.

In agreement with the bulk boundary correspondence generalization to the nonhermitian case, zero modes on $(-\infty,\infty)$ appear when the left and right topological invariants differ $W_L\neq W_R$ or, equivalently, when the local topological invariant $W(x)$ changes sign an odd number of times (if $|W_L-W_R|=1$) or an even number of times (if $|W_L-W_R|=2$) on the interval $(-\infty,\infty)$.
The points $x_i^*$ where the local topological invariant changes must be $|x_i^*|\lesssim w$, since the fields become uniform at larger distances, whereby the local topological invariant becomes uniform as well.
One can infer from the bulk-boundary correspondence and its nonhermitian generalization that, in this case, the zero modes will be localized in close proximity to these points $x\approx x_i^*$. 
Analogously, zero modes on $[0,\infty)$ appear when the topological invariant on the right $W_{R}=W(x\to\infty)$ is nonzero.
If the local topological invariant $W(x)$ is constant, then the zero modes will localize at $x=0$.
However, if the local topological invariant $W(x)$ changes at the points $x_i^*$, then the zero modes will localize in proximity to the points $x\approx x_i^*$.

On the other hand, the quantity ${K}_{L,R}$ distinguishes between exponentially damped oscillations (${K}_{L,R}\neq0$) and exponential decay without oscillations (${K}_{L,R}=0$) at large distances.
Similarly, the value of the function ${K}(x)$ will indicate the presence of oscillations for ${K}(x)\neq0$ and the absence of oscillations for ${K}(x)=0$ at any given point $x$.

\subsection{Examples}

\begin{figure*}[tbp]
 \centering
 \includegraphics[width=\textwidth]{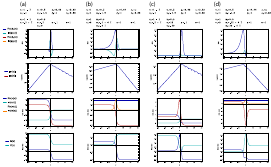} 
\caption{
Zero modes with short hair ($w<\xi$) on $(-\infty,\infty)$ corresponding to a smooth domain wall given by S-shaped, symmetric or asymmetric Pöschl–Teller, or constant fields ${m}(x)$ and ${v}(x)$.
Panels on different rows from top to bottom show the spatial dependence of 
i) the real and imaginary parts of the wavefunctions,
ii) the natural logarithm of the norm of the wavefunctions,
iii) the real and imaginary parts of the fields ${m}(x)$ and ${v}(x)$,
iv) the quantities ${M}(x)$ and ${K}(x)$.
Panels on different columns from left to right correspond to different values of ${m}_{0,1,2}$ and ${v}_{1,2}$, giving
(a) single mode with exponentially damped oscillations on the right and exponential decay on the left for an S-shaped field ${m}(x)$ and a complex S-shaped field ${v}(x)$ with $\Im({v}_L)\neq0$ (nonhermitian),
(b) single mode with exponentially damped oscillations and nonuniform complex phase on the right and exponential decay on the left for a complex S-shaped field ${m}(x)$ with a nonzero imaginary component (nonhermitian).
(c) single mode with exponentially damped oscillations and nonuniform complex phase on the right and exponential decay on the left for an asymmetric Pöschl–Teller field ${m}(x)$ and a complex S-shaped field ${v}(x)$ with $\Im({v}_L)\neq0$ (nonhermitian),
(d) single mode with exponentially damped oscillations and nonuniform complex phase on the right and exponential decay on the left for a complex asymmetric Pöschl–Teller field ${m}(x)$ with a nonzero imaginary component (nonhermitian).
}
 \label{fig:infinf1}
\end{figure*}

\begin{figure*}[tbp]
 \centering
 \includegraphics[width=\textwidth]{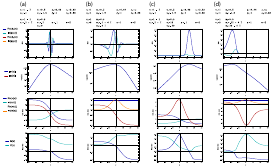} 
 \caption{
Zero modes with long hair ($w>\xi$) on the interval $(-\infty,\infty)$ with S-shaped, symmetric, or asymmetric Pöschl–Teller, or constant fields ${m}(x)$ and ${v}(x)$ as in \cref{fig:infinf1}.
}
 \label{fig:infinf2}
\end{figure*}

\begin{figure*}[tbp]
 \centering
 \includegraphics[width=\textwidth]{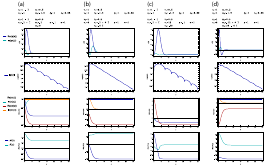} 
\caption{
Zero modes with short hair ($w<\xi$) on the interval $[0,\infty)$ with a smooth domain wall given by S-shaped, symmetric or asymmetric Pöschl–Teller, or constant fields ${m}(x)$ and ${v}(x)$,.
Panels on different rows are as in \cref{fig:infinf1}.
Panels on different columns from left to right correspond to different values of ${m}_{0,1,2}$ and ${v}_{1,2}$ analogous to \cref{fig:infinf1}, giving
(a) single mode with exponentially damped oscillations with nonuniform complex phase for an S-shaped fields ${m}(x)$ and ${v}(x)$ with $\Im({v}_R)\neq0$ (nonhermitian),
(b) single mode with exponentially damped oscillations with nonuniform complex phase for an S-shaped field ${m}(x)$ with a nonzero imaginary component (nonhermitian).
(c) single mode with exponentially damped oscillations with nonuniform complex phase for an asymmetric Pöschl–Teller field ${m}(x)$ and S-shaped field ${v}(x)$ with $\Im({v}_R)\neq0$ (nonhermitian),
(d) single mode with exponentially damped oscillations with nonuniform complex phase for an asymmetric Pöschl–Teller field ${m}(x)$ with a nonzero imaginary component (nonhermitian).
}
 \label{fig:0inf1}
\end{figure*}

\begin{figure*}[tbp]
 \centering
 \includegraphics[width=\textwidth]{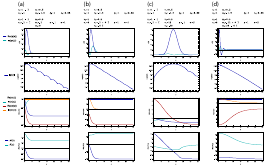} 
 \caption{
Zero modes with long hair ($w>\xi$) on the interval $[0,\infty)$ with S-shaped, symmetric or asymmetric Pöschl–Teller, or constant fields ${m}(x)$ and ${v}(x)$ as in \cref{fig:0inf1}.
}
 \label{fig:0inf2}
\end{figure*}

In this Section, we show some examples of zero modes in the nonhermitian case in  \cref{eq:2generalsolutionsMAIN,eq:2generalsolutionshyperMAIN}.
\Cref{fig:infinf1} shows nonfeatureless zero modes with "short hair" on $(-\infty,\infty)$ with decay lengths $\xi$ larger than the smooth domain wall width $w$.
\Cref{fig:infinf2} shows instead nonfeatureless zero modes with "long hair" with decay lengths $\xi$ smaller than the smooth domain wall width $w$.
\Cref{fig:0inf1,fig:0inf2} shows zero modes on the interval $[0,\infty)$.
The local features are more visible in the long hair regime.
For instance, the zero modes on \cref{fig:infinf2}(c) and \cref{fig:infinf2}(d) and on \cref{fig:0inf2}(d) do not localize within the smooth the domain wall but in proximity to a point $|x_1^*|>\xi$ where the local topological invariant $W(x)$ and the local topological mass ${M}(x)$ change.

\subsection{Experimental detection}

From the definitions of the decay rates and momenta one has that $\mu_{L}^\pm+\mathrm{i} \kappa_{L}^\pm=-s{v}_L\pm\sqrt{{v}_{L}^2+{m}_{L}}$ and $\mu_{R}^\pm+\mathrm{i} \kappa_{R}^\pm=+s{v}_R\pm\sqrt{{v}_{R}^2+{m}_{R}}$, which yield
\begin{equation}\label{eq:EXP1}
\mu_{L,R}+\mathrm{i} \kappa_{L,R}
=\pm{v}_{L,R}\pm\sqrt{{v}_{L,R}^2+{m}_{L,R}},
\end{equation}
which is valid for both hermitian and nonhermitian cases, and both featureless and nonfeatureless zero modes.
In the hermitian or in the asymptotically hermitian cases (i.e., if ${v}_{L,R},{m}_{L,R}\in\mathbb{R}$) and in the regime of exponentially damped oscillations $\kappa_{L,R}\neq0$, one gets $\mu_{L,R}=\pm {v}_{L,R}$ and consequently
\begin{equation}\label{eq:EXP2}
\kappa_{L,R}^2+\mu_{L,R}^2
=-{m}_{L,R},
\end{equation}
for both featureless and nonfeatureless modes.
Note that the left-hand side of \cref{eq:EXP1,eq:EXP2} contains quantities which characterize the wavefunction at large distances, while the right-hand side contains quantities which depend on the "bulk" properties of the system.
Since all these quantities are measurable, one can design an experimental protocol where the bulk dispersion at large distances and the decay rate and momenta of the zero modes are measured independently.
We stress the fact that this experimental protocol requires measuring the  probability density (and wavefunction) of the eigenmodes and not their mere presence, as in conventional tunneling spectroscopies.  
Strong evidence of zero modes is obtained when these quantities satisfy \cref{eq:EXP1} on an extended range of external parameters.

\subsection{Topological phases and differential order}

We note that a general statement can be made on the maximum number of topologically protected modes and the maximum number of topologically distinct phases only by looking at the differential order of the Hamiltonian.
The order of the differential equation determined by the Hamiltonian coincides with the maximum number of possible topologically protected modes and with twice the maximum possible value of the topological invariant as a consequence of the bulk-boundary correspondence.
A second-order differential equation can support at most 2 linearly independent solutions satisfying the boundary conditions.
Therefore, any Hamiltonian that can be reduced to a scalar second-order differential equation as in the modified Jackiw-Rebbi equation in \cref{eq:diffeq} may exhibit at most 2 topologically protected edge modes.
In this case, the critical lines between topologically distinct phases, where the edge modes localize, can only separate phases with topological invariants $W,W'$ with $|W-W'|\le2$, due to the bulk-boundary correspondence.
Hence, the topological invariant must be $|W|\le1$, with a maximum number of 2 nontrivial phases, regardless of the presence of antiunitary and unitary symmetries.
More generally, given that an $n$-order differential equation can support at most $n$ linearly independent solutions satisfying the boundary conditions, any Hamiltonian that can be reduced to a scalar $n$-order differential equation may exhibit at most $n$ topologically-protected edge modes, 
and the critical lines can only separate distinct phases with topological invariants $W,W'$ with $|W-W'|\le n$, which allows the presence of at most $\lfloor n/2 \rfloor$ topologically nontrivial phases with $2|W|\le n$, regardless of the presence of antiunitary and unitary symmetries.

\section{Conclusions}

Concluding, we derived the analytical form of the wavefunctions of nonhermitian topological zero modes localized at smooth domain walls, by extending the modified Jackiw-Rebbi equation to nonhermitian systems with line gaps. 
Using these analytical results, we showed that the nonhermitian bulk-boundary correspondence for these systems can be directly derived by analyzing the asymptotic and localization properties of the zero modes.
We thus identified a universal relation between the decay rates and oscillation wavelengths of the nonhermitian topological zero modes and the underlying scalar fields. 
This relation provides a direct link between the bulk properties and experimentally measurable physical quantities, which are, in principle, accessible via spatially resolved spectroscopies, suggesting an experimental way to probe nonhermitian topological zero modes.
Analogously to their hermitian counterpart, the nonhermitian topological zero modes exhibit different localization behaviors, from featureless zero modes in the sharp domain wall limit to modes with "short" or "long hair" in the presence of smooth domain walls. 
Additionally, we showed that these modes can exhibit either pure exponential decay or exponentially damped oscillations in different regimes.
These findings allow us to understand the physical properties of the zero modes, such as localization and oscillating behavior, that cannot be understood by topological considerations alone.
Notice also that our derivation of the existence of zero-energy modes in the nonhermitian regime (and in the hermitian regime as a special case) does not rely on topological considerations, but on the analysis of the asymptotic behavior of the solutions and, in particular, on their decay rates.
Hence, our approach is inherently more general since it also applies to regimes where bulk topological invariants are ill-defined, such as in the case where the rate of variation of the fields is comparable to the localization length of the modes.
Indeed, previous works on the bulk-boundary correspondence in nonhermitian systems dealt with energy modes localized at sharp domain walls between topologically inequivalent phases~\cite{shen_topological_2018,gong_topological_2018,kawabata_symmetry_2019}. 
Here, we go beyond this simplifying assumption and provide an analytical and formal approach to deal with the more general and realistic case of smooth domain walls. 
Obviously, the sharp domain wall regime can be obtained as a special case of our analysis. 
In this sense, our work generalizes the bulk-boundary correspondence to smooth domain walls.

Finally, we remark that these findings apply to a large range of systems, including topological insulators and superconductors in the hermitian or nonhermitian regimes, and in particular superconductors with $p$-wave pairing~\cite{kitaev_unpaired_2001}, but can be, in principle, extended to superconductors with $s$-wave pairing~\cite{oreg_helical_2010,lutchyn_majorana_2010}, $d$-wave pairing~\cite{tanaka_theory_1995,kashiwaya_tunnelling_2000}, or odd-frequency pairing~\cite{tanaka_symmetry_2012}, and may have relevance for the study of the Dirac equation with domain walls in high-energy physics~\cite{stojkovic_fermionic_2000}.

\begin{acknowledgments} 
The authors thank Masatoshi Sato and Yukio Takaka for their useful suggestions and feedback.
P.~M. is partially supported by the Japan Society for the Promotion of Science (JSPS) Grant-in-Aid for Early-Career Scientists KAKENHI Grant~No.~23K13028 and No.~20K14375, 
Grant-in-Aid for Transformative Research Areas (B) KAKENHI Grant~No.~24H00826,
and
Grant-in-Aid for Transformative Research Areas (A) KAKENHI Grant~No.~22H05111.
\end{acknowledgments}

\appendix

\section{Formal solutions of the modified Jackiw-Rebbi equation\label{app:formal}}

We are specifically interested in the cases where the fields ${m}(x)$ and ${v}(x)$ are approximately uniform for $|x|>w$, and present nonnegligible spatial variations for $|x|<w$.
This setup describes the case of a smooth domain wall between two uniform phases at $x\ll -w$ and $x\gg w$, or a sharp domain wall between two uniform phases at $x<0$ and $x>0$ in the limit $w\to0$. 
In this setup, most of the interesting physics behavior is concentrated near the domain wall, and not at large distances.
In this spirit, we map the whole real line $(-\infty,\infty)$ into the finite segment $(0,1)$ with the substitution 
\begin{equation}
y(x)=\frac12\left(1+\tanh{(x/2w)}\right)=
\frac{{e}^{x/w}}{{e}^{x/w}+1},
\end{equation}
where $w>0$ is a characteristic length describing the spatial variations of fields, i.e., the width of the smooth domain wall localized at the origin $x=0$. 
We thus expand the fields as a power series of $y$, with the assumption that the fields approach their asymptotic values on the left and right exponentially as
\begin{subequations}\label{eq:assumption}
\begin{align}
|{m}(x\to\mp\infty)-{m}_{L,R}|\sim{e}^{- |x|/w},
\\
|{v}(x\to\mp\infty)-{v}_{L,R}|\sim{e}^{- |x|/w}.
\end{align}
\end{subequations}
Here, the characteristic length $w$ is, in principle, arbitrary but will be chosen, in practice, to match the characteristic length scale of the spatial variations of ${m}(x)$ and ${v}(x)$.
The modified Jackiw-Rebbi equation in \cref{eq:diffeq} becomes
\begin{equation}\label{eq:diffeqtanh}
\varphi''(y)
+\frac{1 - 2y+2w s{v}(y)}{y(1-y)}
\varphi'(y)
-\frac{
w^2{m}(y)
}{y^2(1-y)^2} 
\varphi(y)
=0,
\end{equation}
which presents two regular singular points at $y=0,1$.
Assuming a wavefunction in the form 
\begin{equation}\label{eq:solutionsphi}
\varphi(y)=y^{w\alpha_L}(1-y)^{w{\alpha_R}} F(y)
\end{equation}
with $F(y)\neq0$ at $y=0,1$, and expanding asymptotically at the singular points, we obtain the indicial equations for the exponents
\begin{equation}
\label{eq:indicial}
\alpha_L^2+2s{{v}}_L{\alpha_L}-{m}_L
=0,
\qquad
\alpha_R^2 -2s{{v}}_R{\alpha_R} -{m}_R
=0,
\end{equation}
with solutions given by
\begin{subequations}
\label{eq:alphabeta}
\begin{align}
\alpha_L^\pm
&
=
-s{{v}}_L\pm
{q}_L
,
\qquad
{q}_L=\sqrt{{{v}}_L^2+{m}_L},
\\
\qquad
\alpha_R^\pm
&
=
s{{v}}_R\pm
{q}_R
,
\quad\quad
{q}_R=\sqrt{{{v}}_R^2+{m}_R}
.
\end{align}
\end{subequations}
For each set of exponents $({\alpha_L},{\alpha_R})=(\alpha_L^\pm,\alpha_R^\pm)$ the function $F(y)$ is given by the solution of the equation 
\begin{equation}
F''(y)
+
\frac{C_1(y)}{y(1-y)}
F'(y)
+\frac{C_0(y)}{y^2(1-y)^2}
F(y)
=0.
\label{eq:diffeqf}
\end{equation}
where
\begin{subequations}
\begin{align}
C_1(y)=&
{2w{\alpha_L}+1 - 2(w{\alpha_L}+w{\alpha_R}+1)y+2ws{v}(y) }
,
\\
C_0(y)=&
{w^2\alpha_L^2}
-(2w{\alpha_L}+1)(w{\alpha_L}+w{\alpha_R})y
\nonumber\\&
+(w{\alpha_L}+w{\alpha_R})(w{\alpha_L}+w{\alpha_R}+1)y^2
-w^2{m}(y)
\nonumber\\&
+
2w\left(w{\alpha_L}-(w{\alpha_L}+w{\alpha_R})y\right)s{v}(y)
,
\end{align}
\end{subequations}
are functions of $y$.
There are four different choices of the exponents $(\alpha_L^\pm,\alpha_R^\pm)$, and for any of these choices, we obtain a distinct second-order differential equation for the function $F(y)$, which therefore admits two linearly independent solutions $F_{1,2}$.
However, since the modified Jackiw-Rebbi is a differential equation of second order, one can choose at most two solutions that are linearly independent:
The general solution of the equation in \cref{eq:diffeq} is thus given by a linear combination of these two linearly independent solutions, e.g., 
\begin{equation}\label{eq:twoindipendentsolutions}
\varphi_{1,2}^s(y)=y^{{w\alpha_L^\pm}}(1-y)^{{w\alpha_R^\mp}} 
F_{1,2}(y),
\end{equation}
where 
$F_1(y)$ is one of the two solutions of \cref{eq:diffeqf} corresponding to the exponents $(\alpha_L^+,\alpha_R^-)$ and $F_2(y)$ is one of the two solutions corresponding to the exponents $(\alpha_L^-,\alpha_R^+)$, chosen such that $\varphi_{1,2}^s(y)$ are linearly independent and well behaved at $y=0,1$, i.e., $\varphi_{1,2}^s(y\to0)\sim y^{w\alpha_L^\pm}$ and $\varphi_{1,2}^s(y\to1)\sim (1-y)^{w\alpha_R^\mp}$ in order to satisfy the boundary conditions. 
The last condition mandates that the functions converge in absolute value $|F(y)|<\infty$ for $y=0,1$ or, in the worst case, diverge logarithmically for $y\to0$ or $y\to1$.
The solutions $F_{1,2}$ of \cref{eq:diffeqf} can be determined by expanding the function $F(y)$ and the fields ${m}(y)$ and ${v}(y)$ in powers of $y$.
We note again that the choice of the two linearly independent solutions in \cref{eq:twoindipendentsolutions} is not unique.
Furthermore, choosing two linearly independent solutions is possible even if the exponents are degenerate, i.e., when $\alpha_{L,R}^+=\alpha_{L,R}^-$, as long as the functions $F_{1,2}$ are linearly independent.

Transforming back to the real line $(-\infty,\infty)$ the general solution is any linear combination
\begin{equation}
\label{eq:2generalsolutions}
 \varphi^s(x) =
A_1\varphi_1^s(x)+A_2\varphi_2^s(x),
\end{equation}
with $A_{1,2}\in\mathbb{C}$ and
\begin{equation}
\label{eq:2generalsolutions}
 \varphi_{1,2}^s(x) =
 G(\alpha_L^\pm,\alpha_R^\mp,x)
 F_{1,2}(x),
\end{equation}
up to normalization constants with exponents given by \cref{eq:alphabeta} and pseudospin determined by the boundary conditions, where we define $G({\alpha_L},{\alpha_R},x)=y(x)^{w{\alpha_L}}(1-y(x))^{w{\alpha_R}}$ which is thus given by
\begin{align}
 G({\alpha_L},{\alpha_R},x)
 =&
 \frac{{e}^{{\alpha_L \, x}}} {\left({e}^{{x}/{w}}+1\right)^{w{\alpha_L}+w{\alpha_R} }}
 \nonumber\\
 =&
 \frac{{e}^{-{\alpha_R \, x}}} {\left({e}^{-{x}/{w}}+1\right)^{w{\alpha_L}+w{\alpha_R} }}
 \nonumber\\
 =&
 \frac{{e}^{({ \alpha_L}-{\alpha_R})x/2}}
 {\left(2\cosh{(x/2w)}\right)^{w{\alpha_L}+w{\alpha_R}}}
 ,
\label{eq:Gfun}
\end{align}
where we note that $G({\alpha_L},{\alpha_R},x)=G(-{\alpha_L},-{\alpha_R},x)^{-1}$, and with asymptotic behavior at large $x/w\to\pm\infty$ given by
\begin{equation}
\label{eq:Gasymptote}
 G({\alpha_L},{\alpha_R},x\to\mp\infty)
 =
 {{e}^{\mp{\alpha_{L,R} \, x}}}.
\end{equation}
Moreover, if $0<|F_{1,2}|<\infty$, this also mandates that
\begin{subequations}\label{eq:decay}
\begin{align}
\varphi_{1,2}^s(x\to-\infty)&\sim{e}^{+ \alpha_L^\pm \, x}={e}^{+ \mu_L^\pm \, x}{e}^{+ \mathrm{i}\kappa_L^\pm \, x},
\\
\varphi_{1,2}^s(x\to+\infty)&\sim{e}^{- \alpha_R^\mp \, x}={e}^{- \mu_R^\mp \, x}{e}^{- \mathrm{i}\kappa_R^\mp \, x}.
\end{align}
\end{subequations}
where $\mu_{L,R}^\pm=\Re(\alpha_{L,R}^\pm)$ and $\kappa_{L,R}^\pm=\Im(\alpha_{L,R}^\pm)$.

The asymptotic behavior of the functions $G({\alpha_L},{\alpha_R},x\to\pm\infty)$ does not depend on $w$ and on the functional form of the fields ${m}(x)$ and ${v}(x)$, but only on their values ${m}_{L,R}$ and ${v}_{L,R}$ at $x\to\pm\infty$.
We will show that this property also holds for the solutions $\varphi_{1,2}^s(x)$ of the modified Jackiw-Rebbi equation in \cref{eq:diffeq}.

\section{Exact solutions of the modified Jackiw-Rebbi equation\label{app:exact}}

We found that the solutions $\varphi_{1,2}^s(x)$ can be written in closed form in terms of hypergeometric functions in the case where ${m}(y)$ and ${v}(y)$ can be expanded in powers of $y$, and these expansions can be truncated at the second order and first order, respectively, as 
\begin{subequations}\label{eq:expansion}
\begin{align}
{m}(x)=&{m}_0+{m}_1 y(x)+{m}_2 y(x)^2,
\\
{v}(x)=&{v}_0+{v}_1 y(x), 
\end{align}
\end{subequations}
where 
${m}_0={m}_L$, ${m}_1={m}_R-{m}_L -{m}_2 $, ${m}_2= 2 ({m}_L + {m}_R) - 4 {m}(0)$,
${v}_0={v}_L$, and ${v}_1={v}_R-{v}_L$. 
At the zeroth order, \cref{eq:expansion} describes the cases of constant fields 
${m}(x)={m}_{L,R}={m}_0$ when ${m}_1={m}_2=0$ and
${v}(x)={v}_{L,R}={v}_0$ when ${v}_1=0$.
At the first order, \cref{eq:expansion} describes the case where the fields follow an S-shaped curve interpolating between ${v}_L={v}_0$ and ${v}_R={v}_0+{v}_1$ for $x=\pm\infty$, and interpolating between ${m}_L={m}_0$ and ${m}_R={m}_0+{m}_1$ when ${m}_2=0$.
In particular, a symmetric S-shaped curve interpolating between ${v}_L={v}_0$ and ${v}_R=-{v}_0$ for $x=\pm\infty$ is obtained if ${v}_1=-2{v}_0$.
At the second order, ${m}(x)$ in \cref{eq:expansion} describes the symmetric Pöschl–Teller potential with ${m}_{L,R}={m}_0$ in the case where ${m}_1+{m}_2=0$ and an asymmetric Pöschl–Teller potential, i.e., a superposition of a Pöschl–Teller potential and an S-shaped potential otherwise.

If the expansion in \cref{eq:expansion} holds, \cref{eq:diffeqf} simplifies to 
\begin{equation}
F''(y)
+
\frac{c-(a+b+1) y}{y(1-y)} 
F'(y)
-
\frac{a b}{y(1-y)}
F(y)
=0,
\label{eq:diffeqfhg}
\end{equation}
which is the hypergeometric equation 
with $a,b,c$ depending on the exponents ${\alpha_L},{\alpha_R}$ as
\begin{subequations}
\label{eq:abc}
\begin{align}
a,b&=w{\alpha_L}+w{\alpha_R} -w s{v}_1
+\frac12
\nonumber\\&
\pm
\frac12\sqrt{
\left(
2ws{v}_1
-1\right)^2
+4w^2{m}_2
}
,\\
1-c&=
-2ws{v}_L
-2w{\alpha_L}
,
\\
c-a-b&=
+2ws{v}_R
-2w{\alpha_R}.
\end{align}
\end{subequations}
The explicit expressions for $F_{1,2}$ in \cref{eq:2generalsolutions,eq:twoindipendentsolutions} are obtained by taking two solutions of the hypergeometric equation, i.e, choosing two linearly independent solutions out of the Kummer’s 24 solutions of the hypergeometric equation, which yields
\begin{subequations}
\label{eq:kummel}
\begin{align}
f_{1}(y)&={{\,}_2F_1}\left(a,b, c,y\right)
,
\\
f_{2}(y)&={{\,}_2F_1}\left(a,b, a+b-c+1,1-y\right)
.
\end{align}
\end{subequations}

Notice that, if one considers higher order terms in \cref{eq:expansion}, the resulting equation will not reduce to a hypergeometric equation in the general case.
The equation above admits solutions that can be written in terms of two linearly independent solutions given by hypergeometric functions, although the choice is not unique.
Hence, the explicit expressions for $F_{1,2}$ in \cref{eq:twoindipendentsolutions} are obtained by taking
\begin{subequations}\label{eq:hyper}
\begin{align}
F_1(y)
=&{{\,}_2F_1}(a_1,b_1,c_1,y),
\label{eq:hypera}\\\label{eq:hypeb}
F_2(y)
=&{{\,}_2F_1}(a_2,b_2,c_2,1-y),
\end{align}
\end{subequations}
where $a_1,b_1,c_1$ and $a_2,b_2,c_2$ are obtained from \cref{eq:abc} by substituting $({\alpha_L},{\alpha_R})=(\alpha_L^+,\alpha_R^-)$ and $({\alpha_L},{\alpha_R})=(\alpha_L^-,\alpha_R^+)$, respectively, giving
\begin{subequations}\label{eq:ab1ab2}
\begin{align} 
a_{1,2}&=
\pm w({{q}_L}-{{q}_R})
\nonumber\\&+
\frac12+\frac12
\sqrt{
\left(
2ws{v}_1
-1\right)^2
+4w^2{m}_2
},
\\
b_{1,2}&=
\pm w({{q}_L}-{{q}_R})
\nonumber\\&+
\frac12-
\frac12\sqrt{
\left(
2ws{v}_1
-1\right)^2
+4w^2{m}_2
},
\\
c_{1,2}&=1+2{w{q}_{L,R}},
\end{align}
\end{subequations}
provided that the corresponding solutions of the modified Jackiw-Rebbi equation in \cref{eq:twoindipendentsolutions} are linearly independent and can satisfy the boundary conditions.
We notice that $a_1-a_2=b_1-b_2=c_1-c_2$ and that $b_{1,2}=1-a_{2,1}$.

Since ${q}_L$ and ${q}_R$ cannot be negative real numbers by definition, the hypergeometric functions above are well-defined, i.e., the hypergeometric function converges since $\Re(c_{1,2})\ge1$ and therefore $c_{1,2}\notin\mathbb{Z}_\leq$.
Moreover, to satisfy the boundary conditions, we must ensure that the functions defined above are sufficiently well-behaved at the extremes of the interval $0\le y\le1$.
We have that ${{\,}_2F_1}(a,b,c,0)=1$ which gives $F_1(0)=F_2(1)=1$.
Furthermore, the asymptotic behavior of ${{\,}_2F_1}(a,b,c,y\to1)$ depends on the value of $c-a-b$.
Since, $c_1-a_1-b_1=c_2-1=2{w{q}_R}=2w\sqrt{{{v}}_R^2+{m}_R}$, 
by using the asymptotic properties of the hypergeometric functions, one obtains the following asymptotic behaviors, assuming $c_1-a_1,c_1-b_1\notin\mathbb{Z}_\leq$.
If either $a_1\in\mathbb{Z}_\leq$ or $b_1\in\mathbb{Z}_\leq$, the hypergeometric function reduces to a polynomial and one has
\begin{equation}\label{eq:yasymptoticexceptional}
{{\,}_2F_1}\left(a_1,b_1,c_1,1\right)=
\frac{\Gamma(c_1)\Gamma(c_1-a_1-b_1)}{\Gamma(c_1-b_1)\Gamma(c_1-a_1)}\neq0,
\end{equation}
which is nonzero since we assume $c_1-a_1,c_1-b_1\notin\mathbb{Z}_\leq$.
If $a_1,b_1,c_1-a_1,c_1-b_1\notin\mathbb{Z}_\leq$, then
$F_1(y)$ converges for $y\to1$ to a nonzero value if $\Re({q}_R)>0$ (e.g., ${{v}}_R^2+{m}_R>0$ in the asymptotically hermitian case ${v}_R,{m}_R\in\mathbb{R}$) as
\begin{equation}\label{eq:yasymptoticreal}
F_1(y\to1)
=
\frac{\Gamma(c_1)\Gamma(2w{q}_R)}{\Gamma(c_1-a_1)\Gamma(c_1-b_1)}\neq0,
\end{equation}
it is bounded with oscillating behavior if $\Re({q}_R)=0$ and ${q}_R\neq0$ (e.g., ${{v}}_R^2+{m}_R<0$ in the asymptotically hermitian case ${v}_R,{m}_R\in\mathbb{R}$) as
\begin{align}
F_1(y\to1)
&\sim
(1-y)^{\mathrm{i}\Im(2w{q}_R)}
\frac{\Gamma(c_1)\Gamma(-2w{q}_R)}{\Gamma(a_1)\Gamma(b_1)}
\nonumber\\&+
\frac{\Gamma(c_1)\Gamma(2w{q}_R)}{\Gamma(c_1-a_1)\Gamma(c_1-b_1)},
\label{eq:yasymptoticcomplex}
\end{align}
and it diverges logarithmically if ${q}_R=0$ (i.e., ${{v}}_R^2+{m}_R=0$) as
\begin{equation}\label{eq:yasymptoticlog}
F_1(y\to1)\sim
-
\frac{\Gamma(c_1)}{\Gamma(a_1)\Gamma(b_1)}
{\log\left(1-y\right)}.
\end{equation}
Analogously, if $c_2-a_2,c_2-b_2\notin\mathbb{Z}_\leq$ and either $a_1\in\mathbb{Z}_\leq$ or $b_1\in\mathbb{Z}_\leq$, then 
$F_2(0)$ is finite and nonzero, while if $a_1,b_1,c_2-a_2,c_2-b_2\notin\mathbb{Z}_\leq$ then
$F_2(y)$ converges for $y\to0$ to a nonzero value if 
$\Re({q}_L)>0$ (e.g., ${{v}}_L^2+{m}_L>0$ in the asymptotically hermitian case ${v}_L,{m}_L\in\mathbb{R}$),
it is bounded with oscillating behavior if 
$\Re({q}_L)=0$ and ${q}_L\neq0$ (e.g., ${{v}}_L^2+{m}_L<0$ in the asymptotically hermitian case ${v}_L,{m}_L\in\mathbb{R}$),
and it diverges logarithmically if ${q}_L=0$ (i.e., ${{v}}_L^2+{m}_L=0$) with asymptotic behavior given by analogous equations as in \cref{eq:yasymptoticexceptional,eq:yasymptoticreal,eq:yasymptoticcomplex,eq:yasymptoticlog} where $a_1,b_1,c_1\to a_2,b_2,c_2$ and ${q}_R\to{q}_L$.

Hence, if $c_{1,2}-a_{1,2},c_{1,2}-b_{1,2}\notin\mathbb{Z}_\leq$, then
$0<|F_{1,2}(y)|<\infty$ on the whole interval $0\le y\le1$ or diverges logarithmically at the extremes of the interval.
This ensures that $|\varphi_{1,2}^s(y\to0)|\sim y^{w\Re(\alpha_L^\pm)}$ and $|\varphi_{1,2}^s(y\to1)|\sim (1-y)^{w\Re(\alpha_R^\mp)}$.
Finally, we notice that the choice of the functions $F_{1,2}$ in \cref{eq:hyper} ensures that the solutions of the modified Jackiw-Rebbi equation in \cref{eq:twoindipendentsolutions} are linearly independent, since the Wronskian of the two solutions is 
\begin{gather}
W\{\varphi_1^s\left(x\right),\varphi_2^s\left(x\right)\}=
- \frac{\Gamma\left(c_1\right)\Gamma\left(c_2\right)}{\Gamma\left(c_{1}-a_{1}\right)\Gamma\left(c_{1}-b_{1}\right)}
\nonumber\\\times
y^{-sw{v}_L-1}(1-y)^{sw{v}_R-1}
\end{gather}
which is always nonzero since we assume that $c_{1,2}-a_{1,2},c_{1,2}-b_{1,2}\notin\mathbb{Z}_\leq$.

Transforming back to the real line $(-\infty,\infty)$, the functions $F_{1,2}$ in \cref{eq:2generalsolutions} are given by
\begin{equation}
\label{eq:2generalsolutionshyper}
F_{1,2}(x)=
 {{{\,}_2F_1}\left(a_{1,2},b_{1,2},c_{1,2},\frac{1}{{e}^{\mp \, x/w}+1}\right)},
\end{equation}
where we always assume that $c_{1,2}-a_{1,2},c_{1,2}-b_{1,2}\notin\mathbb{Z}_\leq$.
Hereafter, for the sake of simplicity, we will not address explicitly the special cases where either $c_{1}-a_{1}=c_{2}-a_{2}\in\mathbb{Z}_\leq$ or $c_{1}-b_{1}=c_{2}-b_{2}\in\mathbb{Z}_\leq$, which may be obtained as limiting cases.

One has that $F_{1,2}(x\to\mp\infty)=1$.
Furthermore, if $c_{1,2}-a_{1,2},c_{1,2}-b_{1,2}\notin\mathbb{Z}_\leq$, then
$F_1(x)$ converges for $x\to \infty$ to a nonzero value if $\Re({q}_R)>0$ (e.g., ${{v}}_R^2+{m}_R>0$ in the asymptotically hermitian case ${v}_R,{m}_R\in\mathbb{R}$)
while $F_2(x)$ converges for $x\to-\infty$ to a nonzero value if $\Re({q}_L)>0$ (e.g., ${{v}}_L^2+{m}_L>0$ in the asymptotically hermitian case ${v}_L,{m}_L\in\mathbb{R}$);
$F_1(x)$ is bounded with oscillating behavior if $\Re({q}_R)=0$ and ${q}_R\neq0$ (e.g., ${{v}}_R^2+{m}_R<0$ in the asymptotically hermitian case ${v}_R,{m}_R\in\mathbb{R}$)
while $F_2(x)$ is bounded with oscillating behavior if $\Re({q}_L)=0$ and ${q}_L\neq0$ (e.g., ${{v}}_L^2+{m}_L<0$ in the asymptotically hermitian case ${v}_L,{m}_L\in\mathbb{R}$);
$F_1(x)$ diverges polynomially if ${q}_R=0$ (i.e., ${{v}}_{R}^2+{m}_{R}=0$) while
$F_2(x)$ diverges polynomially if ${q}_L=0$ (i.e., ${{v}}_{L}^2+{m}_{L}=0$).
These asymptotic behaviors are summarized in \cref{tab:JRasymptotics}.
It follows that, if $c_{1,2}-a_{1,2},c_{1,2}-b_{1,2}\notin\mathbb{Z}_\leq$, then
$0<|F_{1,2}(x)|<\infty$ on the whole real line or diverges polynomially for $x\to\pm\infty$.

As previously observed, in the hermitian case $\Im({m}(x))=\Im({v}(x))=0$, then the modified Jackiw-Rebbi equation in \cref{eq:diffeq} is real, and hence the solutions $\varphi_{1,2}^s(x)$ are either real or can be linearly combined into a set of linearly independent solutions which are real.
In the nonhermitian case when $\Im({m}(x)\neq0$ or $\Im({v}(x))\neq0$ instead, the solutions $\varphi_{1,2}^s(x)$ cannot be linearly combined into a set of linearly independent solutions which are real.

In Table I of Ref.~\cite{marra_topological_2024} are shown some special cases where the analytical expression of the general solution simplifies.
In particular, we consider the cases where ${v}(x)={v}_{L,R}={v}$ is constant or a symmetric S-shaped curve (${v}_0={v}$ and ${v}_1=-2{v}$) with ${v}_{L,R}=\pm{v}$, and the cases where ${m}(x)={m}$ is constant or a symmetric Pöschl–Teller potential (${m}_0={m}$ and ${m}_1+{m}_2=0$), with ${m}_{L,R}={m}$, and more generally the case where ${q}_L={q}_R={q}$.
In the cases considered, we have $|{v}_{L,R}|={v}$, and thus, we can assume ${v}\in\mathbb{R}$ up to a gauge transformation.

In particular, if ${q}_{L,R}={q}$, one has that $a_{1,2}=a$ and $b_{1,2}=b$ with $a+b=1$.
In this case, the hypergeometric function can be written in terms of the associated Legendre functions as
\begin{equation}
 {{{\,}_2F_1}\left(a,b,c,\frac{1}{{e}^{\mp \, x/w}+1}\right)}
=\Gamma(c)
{e}^{\pm {q} x}
P^{1-c}_{-a}\left(\tanh\left(\mp\frac{x}{2w}\right)\right).
\end{equation}
The correspondence between the hypergeometric functions and associated Legendre functions is the consequence of the fact that the hypergeometric equation can be transformed into a Legendre's differential equation by mapping the interval $0\le y\le1$ to the interval $-1\le z\le1$ by the substitution $z=1-2y=\tanh(x/2w)$.

\end{document}